%% file: ms.tex
\documentclass[sigconf]{acmart}

\usepackage{algorithm}
\usepackage[noend]{algorithmic}
\usepackage{amsmath}
\usepackage{booktabs}

\usepackage{subfigure}

\setcopyright{none} 

\begin{document}
\title{WebFlow: Scalable and Decentralized Routing for Payment Channel Networks with High Resource Utilization} 
\author{Xiaoxue~Zhang, Shouqian~Shi  and Chen~Qian
	\\
	University of California, Santa Cruz
}

\begin{abstract}
    \input{1-abstract.tex}
\end{abstract}





\maketitle

\input{2-introduction.tex}
\input{3-relatedwork.tex}
\input{4-overview.tex}

\input{5-Position.tex}
\input{6-WF1.0.tex}
\input{7-WF2.0.tex}
\input{8-Privacy.tex}

\input{9-evaluation.tex}

\input{91-Discussion.tex}
\input{92-Conclusion.tex}

\bibliographystyle{ACM-Reference-Format}
\bibliography{ms}

\clearpage
\appendix

\input{10-appendix.tex}

\end{document}

%% file: 1-abstract.tex
Payment channel networks (PCNs) have been designed and utilized to address the scalability challenge and throughput limitation of blockchains. Routing is a core problem of PCNs. An ideal PCN routing method needs to achieve 1) high scalability that can maintain low per-node memory and communication cost for large PCNs,  2) high resource utilization of payment channels, and 3) the privacy of users. However, none of the existing PCN systems consider all these requirements. In this work, we propose WebFlow, a distributed routing solution for PCNs, which only requires each user to maintain localized information and can be used for massive-scale networks with high resource utilization. We make use of two distributed data structures:  multi-hop Delaunay triangulation (MDT) originally proposed for wireless networks and our innovation called distributed Voronoi diagram. We propose new protocols to generate a virtual Euclidean space in order to apply MDT to PCNs and use the distributed Voronoi diagram to enhance routing privacy. We conduct extensive simulations and prototype implementation to further evaluate WebFlow. The results using real and synthetic PCN topologies and transaction traces show that WebFlow can achieve extremely low per-node overhead and a high success rate compared to existing methods.


%% file: 2-introduction.tex
\vspace{-1ex}
\section{Introduction}
Blockchain is a promising solution for decentralized digital ledgers.
Since Bitcoin was invented in 2008~\cite{nakamoto2019bitcoin}, there have been many other payment systems emerging based on blockchains, such as Ripple~\cite{armknecht2015ripple}, Stellar~\cite{Stellar}, and Ethereum~\cite{Ethereum}. 
While blockchains have shown the great success as decentralized digital ledgers, \textit{scalability} remains a huge problem with growing numbers of users and transactions~\cite{croman2016scaling, poon2016bitcoin}. For instance, Bitcoin can only support 10 transactions per second at peak in 2020~\cite{TransRate}. 
In contrast, some widely used centralized payment hubs such as Visa and MasterCard can process more than 65,000 transaction messages per second as of June 30, 2019~\cite{Visa}. 
The reason for such a low throughput is that every node processes all transactions and the consensus is achieved by proof-of-work, a time- and resource-consuming process. 
Whenever a new block arrives, all nodes in the network have to process it and update the state of the blockchain.
Hence using blockchains as a global transaction system for massive users is impractical at this moment. 
There are some improvements in basic blockchain such as Bitcoin-NG~\cite{eyal2016bitcoin} and Monoxide~\cite{wang2019monoxide}. However, their performance is limited by the processing capacity of the nodes and network bandwidth, and thus, cannot be used as large-scale global transaction systems.

\begin{figure}[t]
	\centering
	\includegraphics[width=0.45\textwidth]{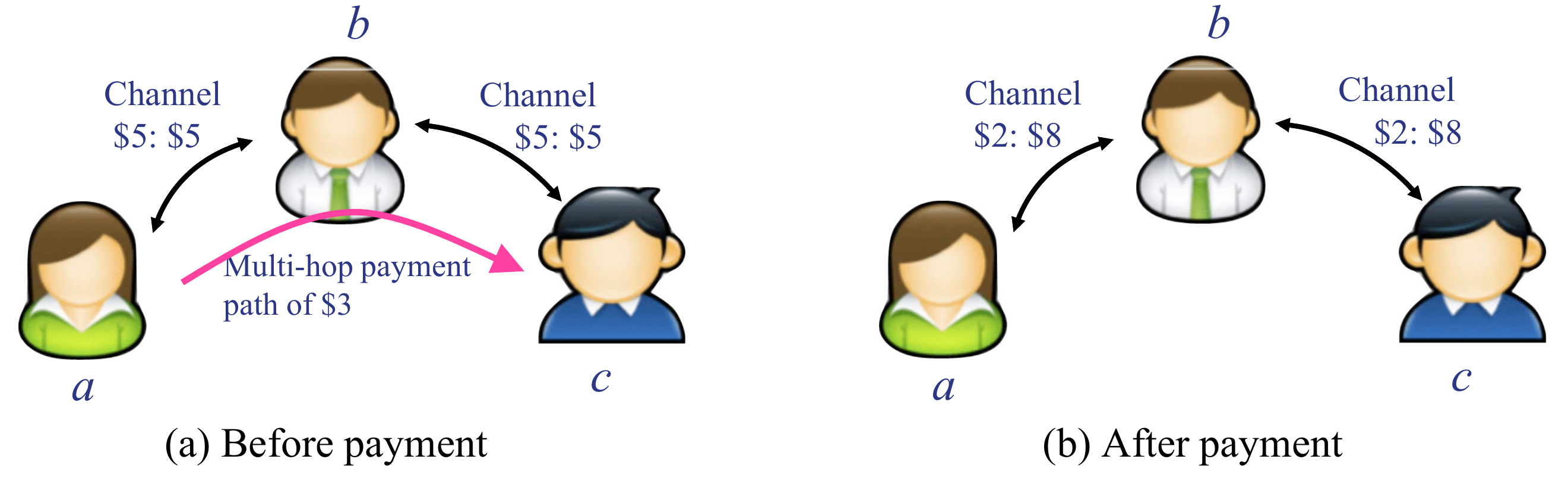}
		\vspace{-2ex}
	\caption{A multi-hop payment in a PCN.}
	\label{fig:channel}
	\vspace{-4ex}
\end{figure}

The recently proposed concept of \textit{payment channel networks} (PCN)~\cite{gilad2017algorand,poon2016bitcoin} 
provides a high-throughput solution for block-chain based payment systems. 
In a PCN, a user $a$ can conduct transactions with another user $b$ through a bi-directional channel.
For this channel, only two transactions need to be recorded on the blockchain: opening and shutting down the channel.  Each user commits a certain fund at the opening of this channel.
Then they can make any number of transactions that update the tentative distribution of the channel's funds as long as the remaining funds allow.
Fig.~\ref{fig:channel}(a) shows a simple example of a PCN. There is a channel $a-b$ between user $a$ and $b$ which includes $\$5$ from $a$ and $\$5$ from $b$, and a channel $b-c$ between user $b$ and $c$ which includes $\$5$ from $b$ and $\$5$ from $c$.
When $a$ pays $b$ $\$3$, the fund distribution in the channle $a-b$ changes to $\$2$ for $a$ and $\$8$ for $b$. 
These transactions only need to be signed by the signatures of $a$ and $b$ but do \textit{not} need to be broadcast to the entire blockchain.
Each user can establish channels to multiple other users. But the channel is not always existed between two arbitrary users, 
and two users sharing a channel usually reflects some level of trust. 
A user can make a transaction with another arbitrary user via a multi-hop path, where any two consecutive users on the path share a channel. 
For example, if $a$ wants to make a payment to another user $c$ without a direct channel, as shown in Fig.~\ref{fig:channel}(a). $b$ has direct channels to both $a$ and $c$. Hence they can use the multi-hop path $a-b-c$ and adjust the fund distribution on the channels $a-b$ and $b-c$ accordingly as in Fig.~\ref{fig:channel}(b). 
The PCN is a promising solution to achieve scalability of blockchains because most transactions can be achieved in an off-chain manner. 
The ultimate vision is that everyone can conduct payment transactions with any other person in the world using a  trust network \cite{ripple2004} without centralized organizations such as banks and governments.


\begin{figure*}[t!]
	\centering
	\begin{tabular}{p{180pt}p{300pt}}
		\centering\includegraphics[width=0.8\columnwidth]{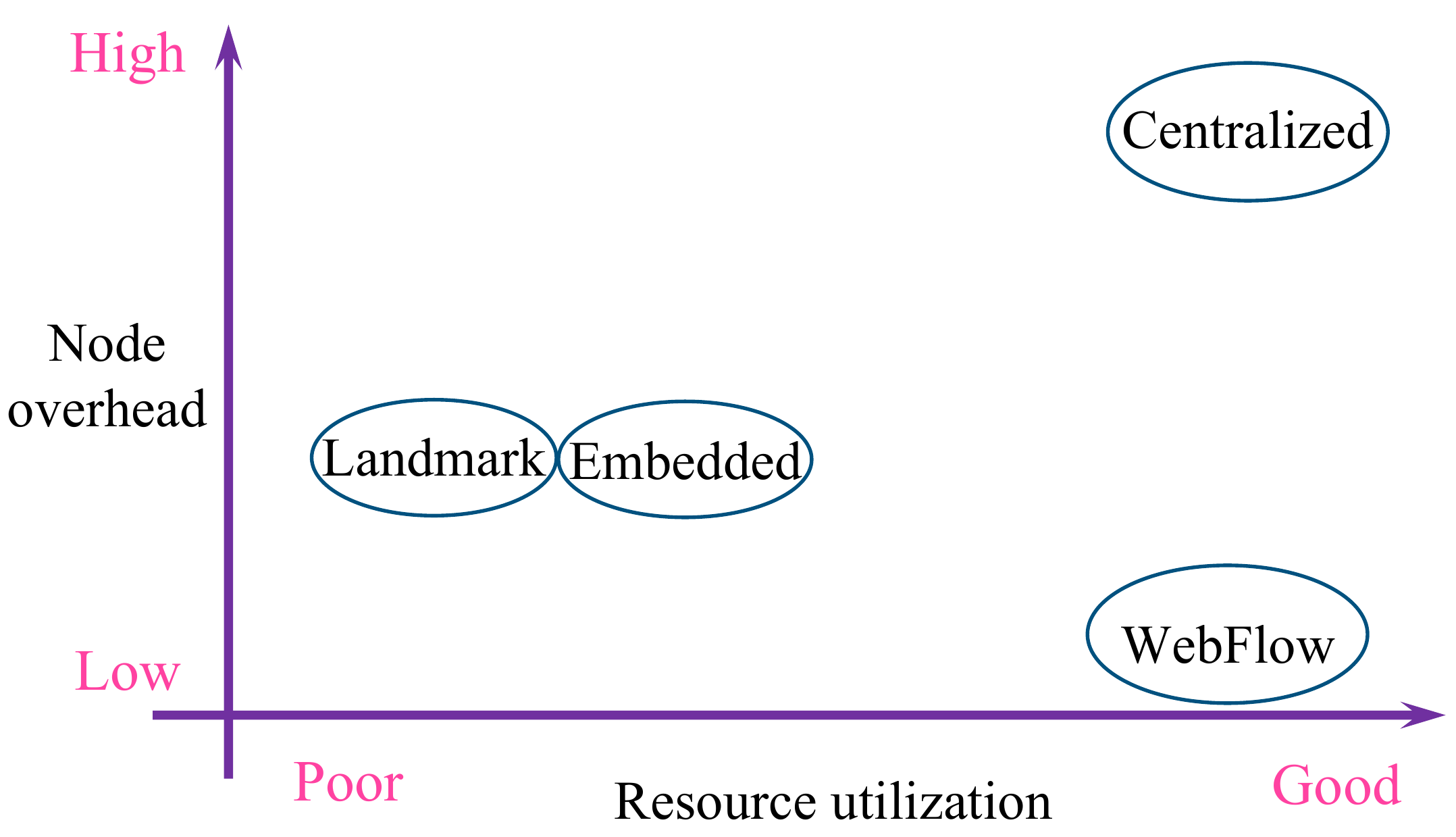}
		\vspace{-4.5ex}
		\caption{Trade-off of PCN routing}
		\label{fig:compare}		
		&
		\centering\includegraphics[width=1.3\columnwidth]{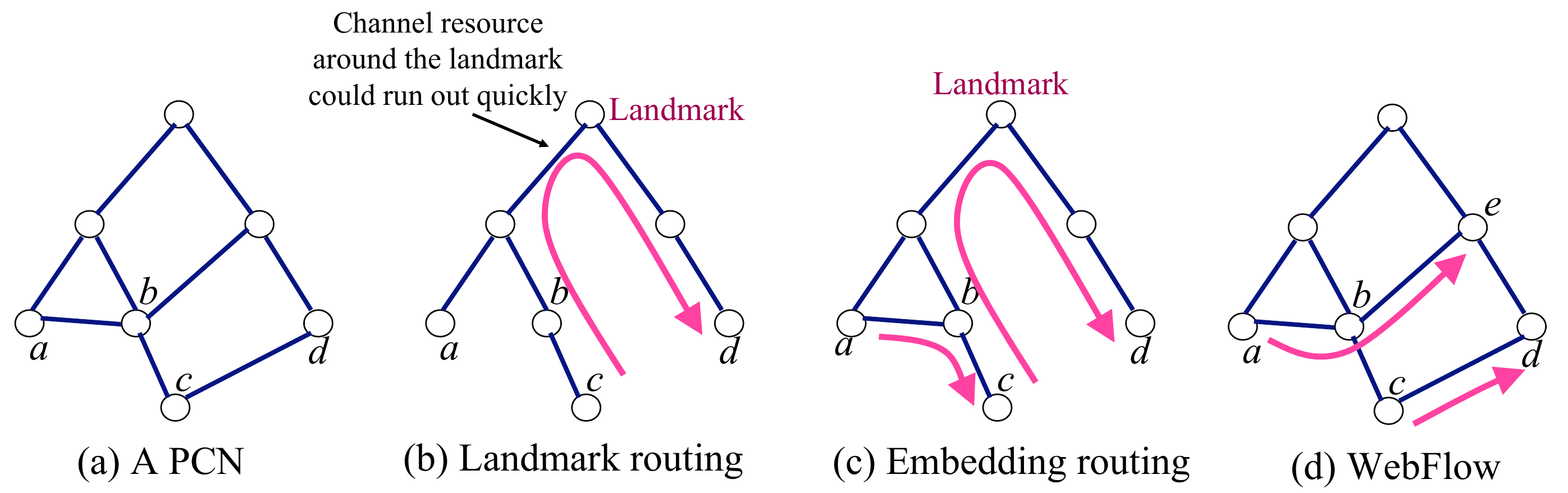}
		\vspace{-4.5ex}
		\caption{Example of different routing methods for PCNs.}
		\label{fig:RoutingMethods}
	\end{tabular}
	\vspace{-5ex}
\end{figure*}


The core problem of a PCN is routing, i.e., finding a path between two arbitrary users. 
Current routing solutions of PCNs can be classified into two types. 1) Centralized routing that assumes every user knows the entire network topology, including all nodes (users), links (channels), and link balances (available funds on each link).  Then each user runs a centralized algorithm to determine the routing paths
 \cite{poon2016bitcoin,wang2019flash,sivaraman2018routing}. 
 \textit{This approach is not a scalable solution} because when a PCN includes many users, each user needs a large memory space to store the network information and every change on any link needs to broadcast to all users. The massive link state updates could cause high network bandwidth cost and the memory cost is a severe problem for \textit{mobile users}.  
 2) Distributed routing for which each user only knows and interacts with a small subset of other users, independent of the entire network size \cite{malavolta2017silentwhispers,roos2017settling}. It is ideal for large-scale PCNs. 
 




Existing distributed routing methods for PCNs, however, have a \textbf{crucial limitation}: they cannot effectively utilize the channel resource in a PCN. We show an example of a PCN with 7 users and 9 channels as in Fig.~\ref{fig:RoutingMethods}(a). One typical distributed routing for PCNs is landmark routing  \cite{malavolta2017silentwhispers}, in which a spanning tree rooted at a landmark user is generated as shown in Fig.~\ref{fig:RoutingMethods}(b). Every transaction (such as the one from $c$ to $d$) will be sent to the landmark first. The landmark knows the whole network topology and will find the path to the destination $d$. This approach does not utilize non-tree channels, and channel resources around the landmark could quickly run out. 
One improvement is embedding-based routing, which is designed to avoid some unnecessary hops in the landmark routing~\cite{roos2017settling}.
It learns a vector embedding for each node. Each node relays each transaction to the neighbor whose embedding is closest to the destination's embedding. Hence it is possible to utilize some non-tree links in a subtree as shown in Fig.~\ref{fig:RoutingMethods}(c). However, many transactions still need to pass through the landmark and cause similar problems. \textbf{Poor channel utilization means fewer transactions can be successfully delivered} in a PCN.

In this paper, we introduce WebFlow, a scalable and distributed routing solution for PCNs with
small per-node overhead and high channel resource utilization. 
As in Fig.~\ref{fig:RoutingMethods}(d), WebFlow allows each node to explore the routing paths without relying on certain ``hot spots'' such as landmarks. Hence, the resource utilization is significantly improved. Meanwhile, each node only stores the information of a few neighbors without knowing the global topology. As shown in Fig.~\ref{fig:compare}, WebFlow provides significantly lower per-node overhead compared to centralized, landmark, and embedding routing while achieving similar channel resource utilization to that of centralized routing. In theory, centralized routing should always achieve better resource utilization than all distributed routing because it can apply any optimization. However, in our evaluation, we compare WebFlow with two recent centralized LCN routing methods  \cite{wang2019flash,sivaraman2018routing} in terms of resource utilization and found it ties with one and outperforms the other. 


WebFlow is a coordinated-based geographic routing protocol for PCNs.
It allows every node to calculate a set of Euclidean coordinates and uses the coordinates to perform coordinated-based greedy routing. We design a system that nodes maintain a multi-hop Delaunay triangulation (MDT) \cite{MDT} based on only the channels with trusted users to achieve a high success ratio of pathfinding. 
To further protect the anonymity of senders and recipients,  our \textbf{important innovation} is to use the property of a distributed Voronoi diagram to achieve the routing tasks. 
 Different from traditional greedy routing, it does not require the coordinates of the destination in a routing message. Instead, it introduces a direction vector to help to determine the path that hides the actual destination.

In summary, this paper makes the following contributions:
\begin{itemize}
  \item We design WebFlow, a new routing protocol for PCNs with low per-node overhead while achieves high resource utilization.
  \item We propose an enhanced version of WebFlow to protect user privacy, i.e., the identities of source and destination of a transaction can be hidden, even if the attacker stands on the path.
  \item We implement WebFlow based on real-world PCN topologies and transactions and build a prototype that can run as a real system. The results show the claimed advantages of WebFlow compared to the state-of-the-art protocols. \vspace{-1ex} 
\end{itemize}

The rest of this paper is organized as follows. Section~\ref{sec:related} introduces the background of PCNs and the related work. The system overview and model are presented in Section~\ref{sec:overview}.
We describe the detail design of the WebFlow protocols in Section~\ref{sec:W1.0} and Section~\ref{sec:W2.0}. Section~\ref{sec:evaluation} presents the evaluation results of WebFlow. Section~\ref{sec:conclusion} summarizes our conclusions. This work does not raise any ethical issues.

%% file: 3-relatedwork.tex
\section{Related work}
\label{sec:related}

PCNs provide a high-throughput solution for blockchains~\cite{gilad2017algorand,poon2016bitcoin}.
To set up a payment channel, two users jointly create a transaction in the blockchain that deposits money for a period of time. Each user commits a certain fund in the opening of this channel. Then the two users can make any number of transactions without broadcasting them to the blockchain.
Two users sharing a channel reflects some level of trust. A user apparently cannot trust arbitrary users. 
In this case, a user can make a payment to (or receive fund from) another untrusted user via a multi-hop path, where any two consecutive users on the path share a channel. The ultimate vision is that everyone can conduct payment transactions with any other person in the world using a  trust network \cite{ripple2004} without the intervention of centralized organizations such as banks and governments.

Routing  is a challenging problem in a PCN.
In Lightning Network \cite{poon2016bitcoin}, each node locally maintains the network topology and a global routing table. In current implementations, users pick the paths by the shortest path and max-flow routing  algorithms.
However, it has three practical problems: 1) The per-node storage cost is $O(|V| + |E|)$ where $|V|$ and $|E|$ denote the number of users and channels in the network respectively; 2) the per-transaction computation cost is $O(|V| \cdot |E|^2)$ using the max-flow routing algorithm; 3) every update of any channel will be broadcast to all nodes. Hence it leads to scalability problems, especially for mobile users. 

Recent work such as Spider \cite{sivaraman2018routing} and Flash \cite{wang2019flash} use centralized routing. 
Spider actively accounts for the cost of channel imbalance by preferring routes that re-balance channels, and it proposes a centralized offline routing algorithm to maximize the success volume of payments. But the centralized scheme still has high probing overhead. Flash reduces the probing overhead by treating elephant and mice payments differently. It requires each user to maintain a routing table for mice payments, and periodically refreshes the routing table when the local network topology is updated. It still applies a huge memory cost for each user.

To reduce the per-node overhead in PCN, landmark routing and embedding-based routing have been proposed as distributed routing. 
In landmark routing, a few users are selected as landmarks, and they are responsible to store routes to all other nodes. Non-landmark nodes only need to store routes to the landmarks.
SilentWhispers \cite{malavolta2017silentwhispers} utilizes landmark routing. It performs a periodic breadth-first search to find the shortest path from the landmarks to the sender and recipient. All paths need to go through the landmarks, which makes the channels around the landmarks over-used and some paths could be unnecessarily long. 
Other channels might be under-used.  
Embedding-based routing is designed to avoid some unnecessary hops in the landmark routing. It learns a vector embedding coordinate for each user based on the structure of spanning trees generated by landmarks. Each user relays transactions to the neighbor whose embedding is closest to the recipient’s embedding. 
SpeedyMurmurs \cite{roos2017settling} uses embedding-based routing. Computing and updating the coordinates as the topology and channel balances change is a major challenge of this approach.

Other overlay network routing methods such as distributed hash tables (DHTs) cannot be used for PCNs because one cannot be forced to build a channel with an arbitrary user. In DHT, A node picks its neighbors according to a certain structure and mapping rules. But these neighbors might not be its direct neighbors in the PCN, and the paths to these neighbors might be long, resulting in a high routing stretch in the PCN.
Kademlia~\cite{maymounkov2002kademlia} enforces a structured topology. However, in reality, as users can set up and shuttle down channels anytime, the topology is unstructured and keeps changing.

There are many existing Ad-Hoc routing protocol such as DSR, AODV and GPSR. However, GPSR only works for planar graphs. In both DSR and AODV, route discovery is based on flooding, which introduces considerable routing overhead, and can not be applied to the large scale PCNs. Besides, the initial balance of a payment channel is deposited by the users during the channel setup, and is kept updating during each transaction. So the route maintenance mechanism in both DSR and AODV may not work well in such dynamic networks.

Compared to existing work, WebFlow is the first solution that considers both the scalability of user overhead and channel resource utilization in PCN routing. 

%% file: 4-overview.tex
\section{Overview}
\label{sec:overview}

\begin{figure*}[t]
	\centering
	\subfigure[Network with random coordinates]{
		\label{fig:random_100}
		\includegraphics[width=0.3\textwidth]{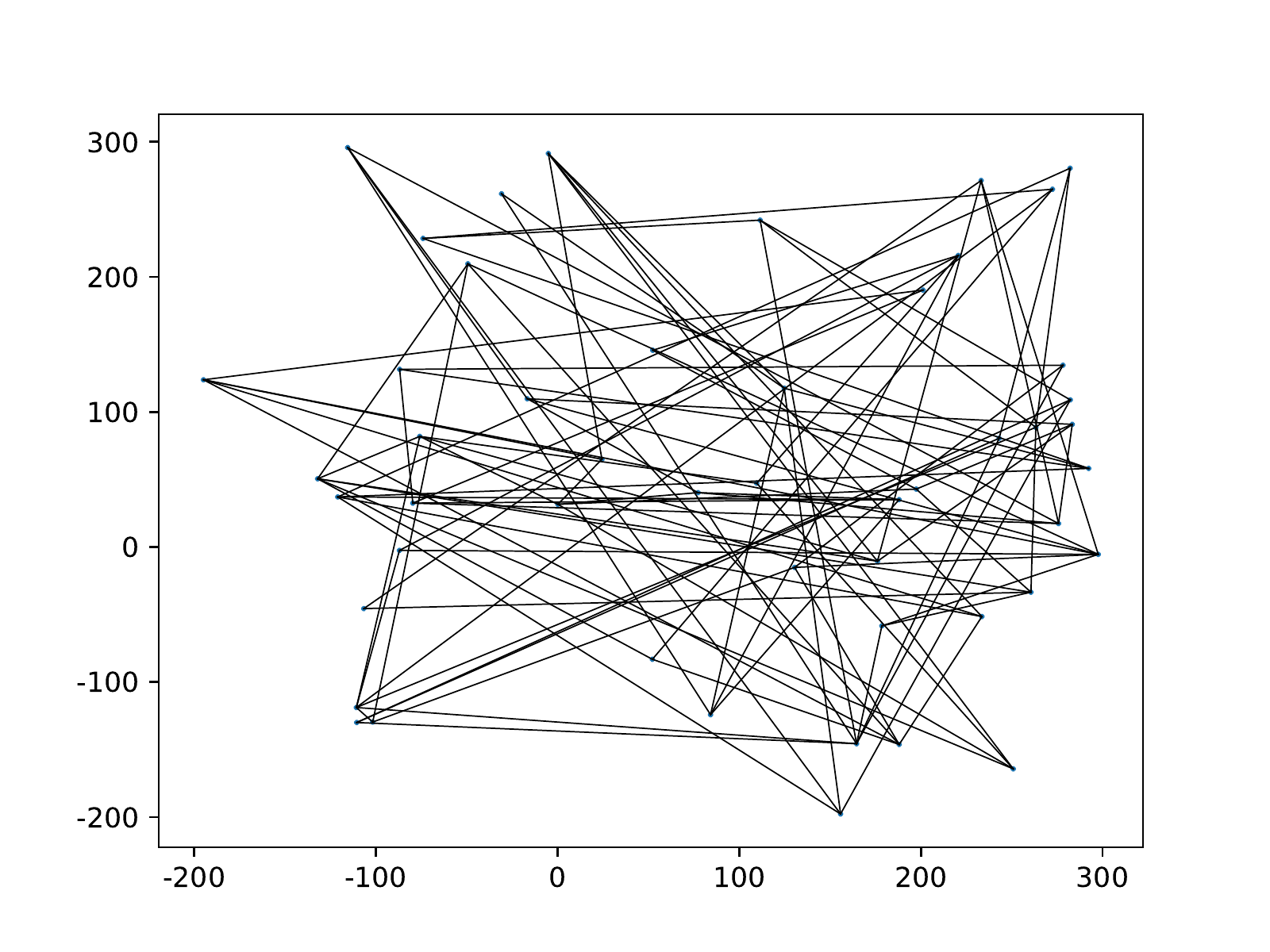}}
	\subfigure[WebFlow coordinates]{
		\label{fig:MDS_100}
		\includegraphics[width=0.3\textwidth]{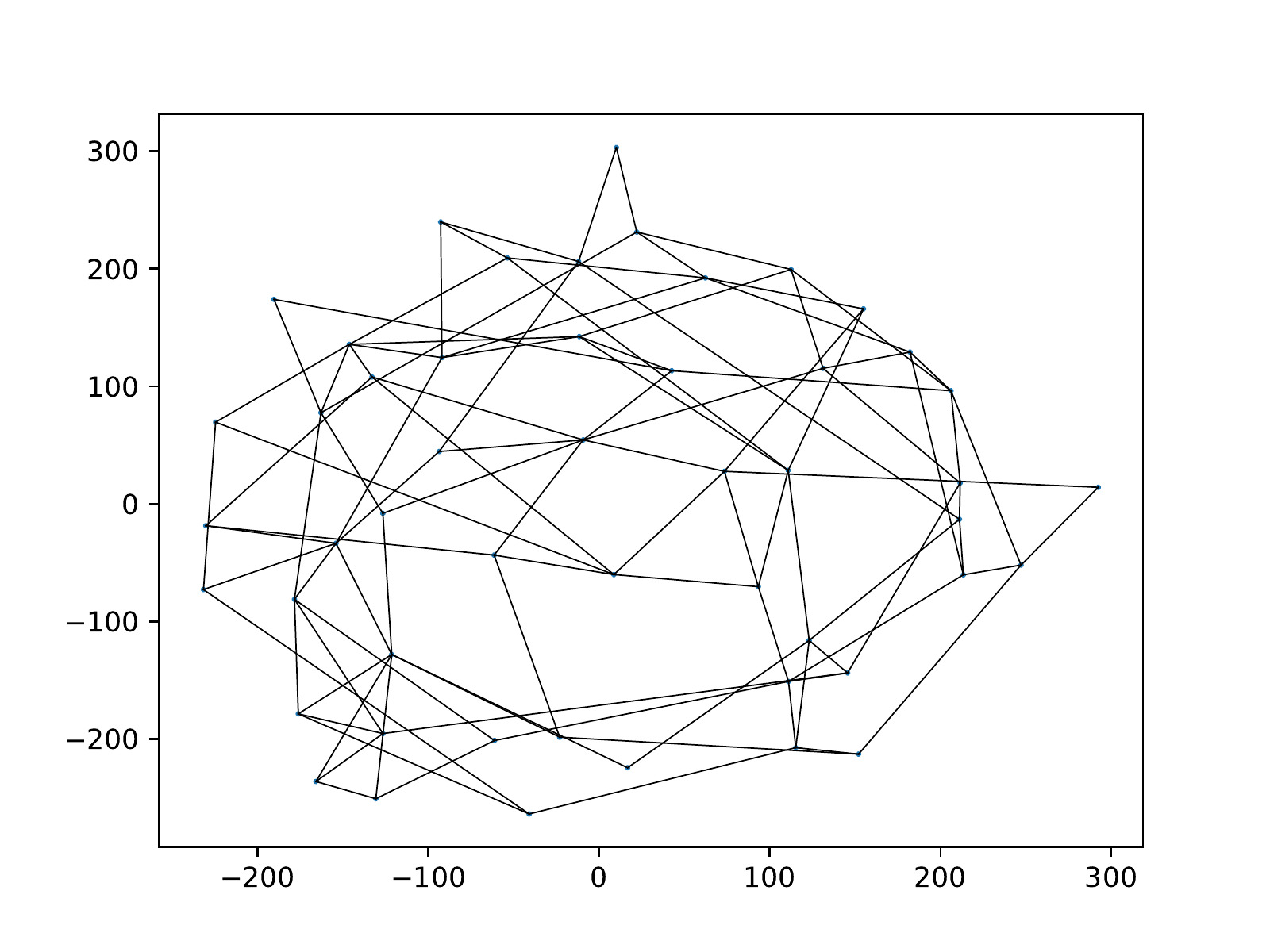}}
	\subfigure[The MDT of the set of nodes]{
		\label{fig:MDT_100}
		\includegraphics[width=0.3\textwidth]{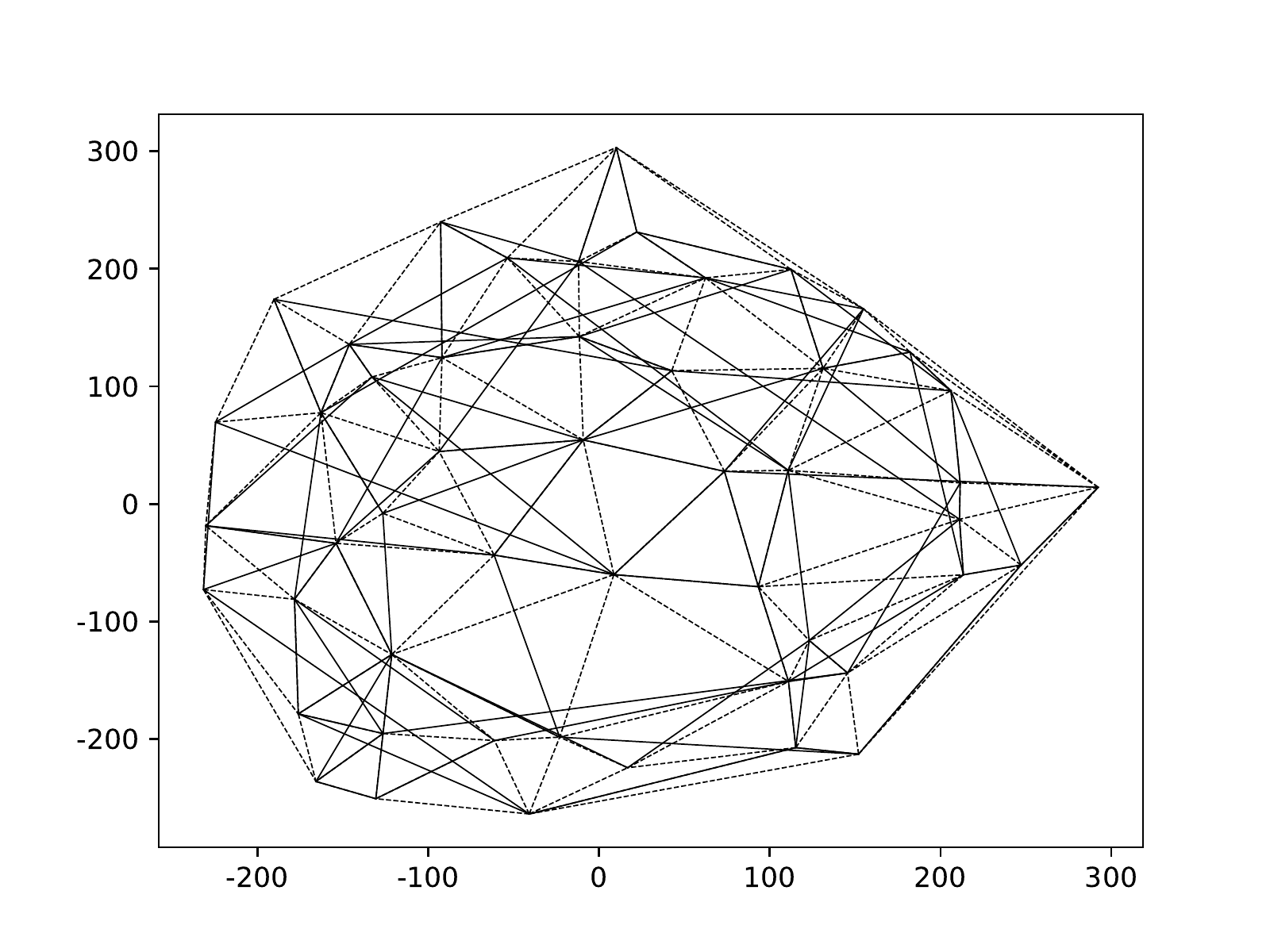}}
		\vspace{-3ex}
	\caption{Original PCN graph, the graph after node positioning, and MDT graph.}
	\label{fig:coordinate}
	\vspace{-2ex}
\end{figure*}

\subsection{Network Model}
WebFlow is a distributed routing protocol for large-scale PCNs. In a PCN, each user is called a \textit{node}. 
The bi-directional payment channel shared by two nodes is called a  \textit{physical channel} or \textit{direct link}, and these two nodes are called \textit{direct neighbors}. 
The fund of one direction of a link is called the balance of that direction.
We model a PCN as a graph $G = (V, E, \Psi)$, where $V$ is the set of nodes, $E$ is the set of links with a weight function $w$, and $\psi_{uv}, \psi_{vu} \in \Psi$ are the balances of link $e=(u,v)$ in two directions. 
Furthermore, a path $p$ is a sequence of links $e_1 ... e_k$ with $e_i = (v_i, v_{i+1})$ for $1 \le i \le k-1$. The parameter $\psi$ describes the amount of funds that can be transferred between two nodes sharing a link. The path of a transaction is accepted only if the amount of this transaction is less than every fund $\psi_1 ... \psi_k$ along this path.
Every node knows the links and their balances to its neighbors. 

\textbf{Problem definition.}
The routing problem of WebFlow is described as follows. Consider a transaction $t$ initiated by \textit{sender} $s$ that should be received by the \textit{recipient} $r$. WebFlow needs to find a path from $s$ to $r$, where two consecutive nodes on the path should share a physical link (payment channel) and each link has enough balance to make the payment to the next-hop. The success of routing implies that $s$ can make a transaction with $r$ by a sequence of transactions involving other nodes, even if $s$ and $r$ have no trusted channel. 
We have three objectives: 1) each node should have limited memory and communication overhead, which is independent of the network size; 2) WebFlow should achieve a high success rate of transaction routing; 3) WebFlow should achieve high channel resource utilization, which can be measured by in the total amount of successfully-processed transaction funds over a period of time.

\subsection{Attacker Model}
\label{sec:attmodel}
We consider the user privacy of their transactions in a fully distributed PCN. Our primary attack scenario is the malicious users interested in other's financial situation. They are a group of honest-but-curious users that passively observes the channels related to them and tries to infer the source and destination of the transactions.
They only know the channels to their neighbors, but cannot access the routing information stored at other honest users. 
The attackers here aim to undermine the user's privacy and profit from it rather than perform a denial-of-service attack. 

Similar to PrivPay \cite{moreno2015privacy} and SpeedyMurmurs \cite{roos2017settling}, our goal is to hide values, and achieve anonymity of sender and receiver when making transactions. We use the term \textit{value privacy}, \textit{sender/receiver anonymity} respectively to refer to these three privacy goals.

\textit{Value privacy:} 
We say that the PCN can achieve value privacy if it is impossible for any adversary to know the total value of a transaction between two honest users.
Let $s$ and $r$ be two non-compromised nodes, and $p$ is the transaction path from $s$ to $r$. We consider two situations: 1) all the nodes along the path are non-compromised, and 2) there exist one or more malicious nodes on the path. In the first case, the attacker cannot obtain any information about the transaction value. In the second case, as a malicious node can monitor how much funds go through itself, it could easily know the transaction value along this path. To prevent this kind of attack, we can just divide the total value into $k$ parts in advance, and select $k$ paths to complete this transaction. As the malicious node can only know the value on the path where it exists, it can't know the total value. Even if the malicious nodes on different paths collaborate, they still do not know how many paths are chosen.

\textit{Sender anonymity:}
We say that a PCN can achieve sender anony-mity if the adversary cannot determine the original sender of a transaction. 

\textit{Receiver anonymity:}
Similarly, a PCN has receiver anonymity if the adversary cannot determine the actual receiver of a transaction. 

We \textit{formally define a metric to evaluate the sender/recipient anony-mity} of a network and its routing algorithm as follows.
The anonymity measure follows the anonymity definition that has been used for anonymous routing such as ~\cite{AnonyMeasurePET02,AnonyMeasurePET02-2,zhuang2005cashmere}. Anonymity is the
state of being not identifiable within a set of subjects, i.e., the \textit{anonymity set}. In a network with $N$ nodes, ideal anonymity is achieved when all nodes look equally likely to be the sender or recipient to an attacker, that is the anonymity set contains all the nodes in the network. However, it is impossible in real-world systems. The attacker can deduce that some nodes are more likely to be the sender or recipient based on information leaked from the system.
So the attacker can assign each node $u$ a probability $p_u$ as being the sender or recipient of a transaction using knowledge of leaked information from the system. All the nodes in the network are denoted as a finite set $V$, where $|V| = N$.
Then the system entropy is defined as:
\begin{equation}
H(V) = -\sum_{u \in V} p_u log_2(p_u)
\end{equation}

If we have ideal anonymity, all nodes look equal to the attacker. We can get $p_u = \frac{1}{N}$ for all the nodes in the system. And the entropy of ideal anonymity is $H_m(V) = \log_2(|V|)$. This is the maximum entropy the system can achieve.

The anonymity of the system is measured by the entropy of this system over the entropy of system ideal anonymity.
The anonymity of the system is measured as:
\begin{equation}
\frac{H(V)}{H_m(V)} = \frac{-\sum_{u \in V} p_u log_2(p_u)}{\log_2(|V|)}
\end{equation}

Again, to measure the anonymity of our system, we need to consider two situations: 1) no attacker on the path, and 2) one or more attackers on the path.  We use $V_A$ to denote the set of attackers, and $P_a$ to denote the path set.
The anonymity of WebFlow will be analyzed in detail in Section~\ref{sec:privacy}. 

\subsection{Analysis methodology of this work}

From our observation of real PCN topologies, they are not any regular graphs such as grids or trees. Hence, it is \textbf{impossible} to use theoretical formulation to analyze the routing performance or anonymity of a routing algorithm. In this paper, we will use \textbf{extensive simulations with both real and synthetic network topologies} to analyze the routing performance or anonymity. In addition, we use prototype implementation to demonstrate that WebFlow can work in practice.  



%% file: 5-Position.tex

\section{Baseline Design of WebFlow}
\label{sec:W1.0}

To achieve distributed and scalable routing, WebFlow utilize the idea of coordinate-based greedy routing that has been widely used for wireless networks \cite{GPSR,MDT}. The basic idea is that in wireless networks, each node knows its geographic coordinates as well as its neighbors'. Then without knowing the whole topology, each node can simply forward a packet to a neighbor who is geographically closest to the destination. The major problem is that a node could be a local minimum and none of its neighbors is closer to the destination than itself. This problem can be solved in various ways  \cite{GPSR,MDT}. Geographic greedy routing is very scalable because each node only needs to know its direct neighbors. 

We would like to apply the idea of coordinate-based greedy routing in WebFlow for PCNs. However, one immediate problem is that many users do not want to share their geographic locations in PCNs which could be a threat to their privacy. In addition, the geographic distances of PCNs do not reflect the routing cost, while in wireless networks distances can be an estimate of routing difficulties. Hence, we propose to use \textbf{virtual coordinates in a Euclidean space} that reflect the PCN topology features. 

Every PCN has multiple web servers as its interface of registration and user interactions. 
The web servers in WebFlow do \textbf{not} participate in routing and only support coordinates computation. Hence it avoids the landmark limitations in landmark-based routing \cite{malavolta2017silentwhispers,roos2017settling}. 
We assume there are no more than $\frac{1}{3}$ malicious servers in our systems. The malicious servers try to stop the progress of the protocol by playing denial-of-service attack, or providing wrong coordinates. To guarantee the correctness of virtual coordinates, the user can simply send coordinate request to all the servers. If it receives same results from more than half of the servers, the user will adopt this virtual coordinate.


\begin{figure*}[t!]
	\centering
	\begin{tabular}{p{180pt}p{300pt}}
		\centering\includegraphics[width=0.5\columnwidth]{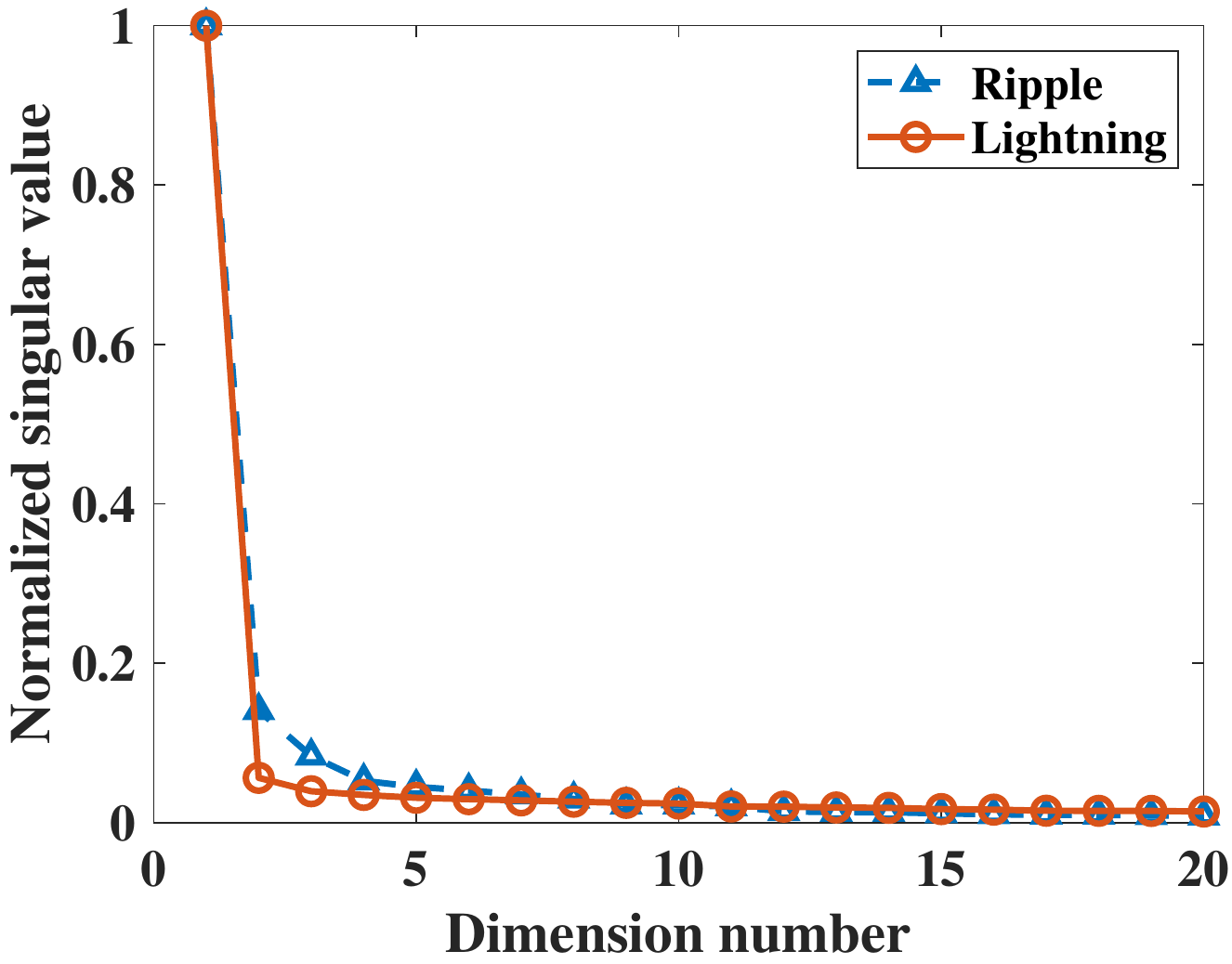}
		\vspace{-2ex}
		\caption{SVD analysis of PCNs}
		\label{fig:svd}		
		&
		\centering\includegraphics[width=1.3\columnwidth]{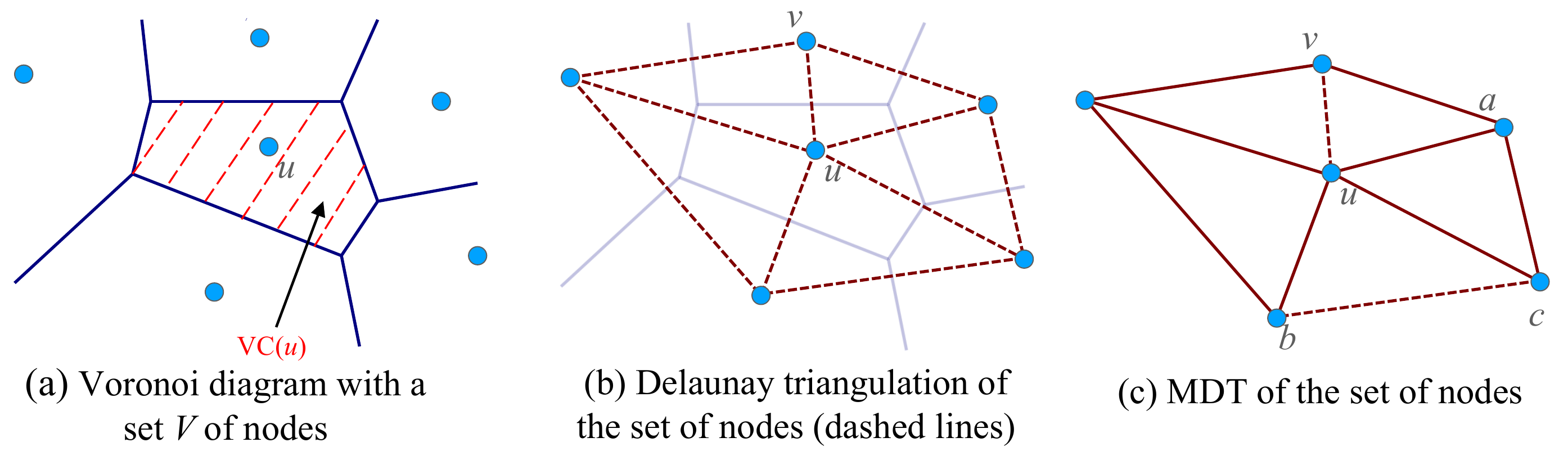}
		\vspace{-4.5ex}
		\caption{Voronoi diagram, Delaunay triangulation (DT), and MDT}
		\label{fig:DT}
	\end{tabular}
	\vspace{-5ex}
\end{figure*}

\vspace{-1ex}

\subsection{Coordinates computation}
For the PCN $G = (V, E, \Psi)$ in a Euclidean space $S$, each node will be assigned a set of coordinates $c^S$ in $S$ by the web servers. The goal is to let nodes maintain coordinates such that two neighbors are relatively close in the Euclidean space. For two arbitrary nodes, their network distance (e.g., in hop-count) would be proportional to their Euclidean distance.  In this way, coordinate-based greedy routing will be more likely to succeed. 
We denote the actual hop-count between nodes $u$ and $v$ as $h_{uv}^S$. The Euclidean distance between the coordinates of $u$ and $v$ is denoted as $d_{uv}^S$.

The web servers randomly select $k$  nodes $T = \{T_1,...,T_k\}$ called \textit{anchors} where $k=d+1$ if we use a $d$-dimensional Euclidean space.  
\footnote{Note that for a $d$-dimensional Euclidean space, $k$ needs to be at least $d+1$\cite{ng2002predicting}.}
For each anchor, the server recursively visits its neighbors, resulting in a spanning tree. Then the server has $k$ spanning trees and knows the hop-counts between all pairs among the anchors, i.e., a $k \times k$ hop-count matrix characterizing the distances between every pair of anchors. 
The server should find a set of coordinates $c_{T_1}^S$,...,$c_{T_k}^S$, such that the Euclidean distance can reflect the hop-counts with minimal errors. We apply an existing algorithm multi-dimensional scaling \cite{MDS} to obtain the coordinates. 
Once the coordinates of anchors, $c_{T_1}^S$,...,$c_{T_k}^S$, are determined, each node contacts the web server to get the coordinates of anchors and their hop-counts to the anchors. Each node can determine its own coordinates by minimizing the overall error between actual hops and computed distances to these anchors. 
Since each user is responsible to compute its own coordinate, the system can be scaled to large sizes. Actually, the most time-consuming part in coordinate computation is initialization when the web servers build a spanning tree for each anchor and compute the coordinates of anchors. 
However, even for a real-world PCN topology generated from Ripple network with 1,870 nodes, it only takes 5 seconds to initialize. And it takes less than 10 ms for each user to generate its own coordinate.

We show an example of a PCN in Fig.~\ref{fig:coordinate}. Fig.~\ref{fig:coordinate}(a) shows the network topology with randomly assigned coordinates. After the WebFlow coordinate assignment in a 2D space, the network is shown in Fig.~\ref{fig:coordinate}(b). We find that the WebFlow coordinates show a better correlation between the Euclidean distances and hop counts. 



\textbf{Choice of dimensionality.} One immediate question is that how many dimensions of the Euclidean space we shall use to characterize the network topologies of PCNs. We use Principal Component Analysis (PCA) to find an appropriate dimensionality. PCA relies on Singular Value Decomposition (SVD). The input of SVD is an $n\times n$ matrix $M$ of the hop-counts of all nodes. SVD factors $M$ into the product of three matrices $M=USV^T$, where $S$ is a diagonal matrix with non-negative elements $s_i$. If singular values $s_1, s_2, ..., s_d$ are much larger than the rest, we may approximate $M$ using $d$-dimensional Euclidean spaces. We analyze two real-world payment channel network topologies: Ripple~\cite{armknecht2015ripple} and Lightning~\cite{poon2016bitcoin}, whose details will be presented in Sec.~\ref{sec:evaluation}.  In Fig.~\ref{fig:svd} we show the singular values of the two PCNs. We find the first three singular values of Ripple and the first two of Lightning are significantly larger than the rest. Hence in WebFlow we will use 3D space and show some comparison with 2D and 4D in evaluation. 

%% file: 6-WF1.0.tex

\vspace{-1.5ex}
\subsection{WebFlow Routing}

One major challenge of greedy routing using virtual coordinates is that it may stuck at a local minimum. 
WebFlow is based on a distributed data structure called Multi-hop Delaunay triangulation (MDT) \cite{MDT}. 
An important feature of MDT is that, each node only needs to maintain the link and path information to a few neighboring nodes, independent of the network size.

For a given set of nodes $V$ in a Euclidean space, the Voronoi diagram is a  partition of space into regions (called cells) such that each cell contains one node and the node is closer to any point in the cell than all other nodes. An example in Fig.~\ref{fig:DT}(a) shows the Voronoi diagram of a 2D space with 6 nodes. For all points in Voronoi cell $VC(u)$,  node $u$ is closer to them than any other node in $V$. A Delaunay triangulation, shown in Fig.~\ref{fig:DT}(b), is a dual graph of the Voronoi diagram, where two nodes in $V$ are connected if their Voronoi cells share an edge. Two nodes $u$ and $v$ are called \textit{DT neighbors} if they are connected in $DT(V)$. 
An important proven feature is that greedy routing on the DT edges (i.e., assuming all DT neighbors are connected) always succeeds to find the destination without encountering a local minimum \cite{MDT}.

However, in reality, not every DT edge is connected. In a PCN two nodes cannot be forced to generate an actual channel. 
As shown in Fig.~\ref{fig:DT}(c), the DT edges $uv$ and $bc$ are not connected. A MDT is a distributed protocol to generate a distributed data structure such that 1) each node knows its DT neighbors; 2) for a DT neighbor without a direct channel to it, MDT provides a multi-hop channel path to the DT neighbor. Hence greedy routing with an MDT can guarantee to find an end-to-end path from a source to the destination, using the destination coordinates. This feature can be extended to $d$-dimension for $d\ge 2$.

We denote $DT(V)$ as a tuple $\{<u,N_u> | u \in V\}$, where $N_u$ is the set of $u$'s DT neighbors, which is locally maintained by $u$. 
The set of $u$'s direct neighbors is denoted as $C_u$. Then $u$ determines its DT neighbor $N_u$ by calculating a local DT of $C_u$. 

The \textbf{major challenge} of applying DT to PCNs is that two DT neighbors may not have a physical channel. 
To address this problem, the MDT is specified as a 3-tuple $\{<u,N_u,F_u> | u \in V\}$, where $F_u$ is a soft-state forwarding table that helps to find multi-hop channel path to the DT neighbors without direct channel. The extension of MDT from DT is that a node in $N_u$ is not necessarily directly connected to node $u$ by physical channels. 
We say that those nodes who are $u$'s DT neighbors but not direct neighbors communicate via a \textit{virtual link} with $u$.
Even if two DT neighbors do not have a channel, they can be connected by virtual link because there is an underlying multi-hop path of real channels that connects one DT neighbor to the other. 
Note that in WebFlow, no node should maintain a global view of the MDT. Each node only maintains local information $<u,N_u,F_u>$ that is independent of the network size. The routing decisions are made locally. In addition, there are no supernodes that handle most payments such as the landmarks in landmark routing. Hence WebFlow is highly scalable and decentralized. \textbf{Given the destination coordinates, WebFlow routing using the MDT graph always finds a path to the destination based on local decisions of the nodes on the path.}

The forwarding table $F_u$ is used by $u$ to help to find the physical path from one node to its DT neighbors. Every entry in $F_u$ is specified by a 4-tuple $<source, pred, succ, dest>$, source and dest being the node and its DT neighbor node in the path respectively, pred and succ being the predecessor and successor nodes of $u$ in this section of the path. 
For an example of a forwarding path, consider the MDT graph in Figure~\ref{fig:DT}(c). Nodes $u$ and $v$ have no channel, but a multi-hop path $u-a-v$ supports the virtual link $uv$. 
So the currency is forwarding along the path $u-a-v$. The forwarding tuples using along the path in the nodes are: $<-,-,a,v>$ in node $u$, $<u,u,v,v>$ in node $a$, and $<-,-,a,u>$ in node $v$. 

The routing algorithms of WebFlow contain several MDT protocols including the forwarding protocol, join protocol, and maintenance protocol. The forwarding protocol determines how a node should locally decide the next-hop node when routing to a receiver in a correct MDT. The other protocols are used to maintain a correct MDT graph. They are all decentralized algorithms.

\vspace{-2ex}
\subsubsection{Forwarding Protocol}
Consider a payment $t$ initiated by node $s$ that should be received by node $r$, and the payment value is $\omega$. For each node $u$ received forwarding request, it first checks whether it is the receiver. If it is, $u$ does not need to forward. Otherwise, if $u$ can find a direct neighbor $v$ closest to the receiver, it will check that if the channel $uv$ has enough capacity to support the payment. If the channel can support the payment, that is $\psi_{uv} > \omega$, $u$ sends the payment to $v$.
Otherwise, $u$ finds the DT neighbor $v'$ that is closest to $r$ among all $u$'s DT neighbors. Then, $u$ needs to probe the virtual link to check if the underlying multi-hop path of this virtual link has enough capacity to support the payment. If it can, $u$ sends the payment to $v'$ using the virtual link. 

If both situations fail due to channel capacity limitation, $u$ traverses all of its direct neighbors to check if there exists a direct neighbor $v$ closer to the receiver with enough channel capacity to support the payment. In other words, the Euclidean distance $d(v,r)<d(u,r)$, and $\psi_{uv} > \omega$. 
If such a direct neighbor does not exist, $u$ selects 5 DT neighbors closer to the receiver. 
Note that $u$'s DT neighbors that are closer to the receiver may be less than 5, in this case, $u$ just selects all these satisfied DT neighbors. Then $u$ probes the virtual links to these selected DT neighbors until it finds a DT neighbor with a virtual link that can support the payment. If all these selected DT neighbors fail to fulfill the payment, we assume that this payment is failed.
Note that if $u$ selects too many DT neighbors to probe, it will introduce large probing overhead and communication cost. If $u$ only probes small number of DT neighbors, it will lower the success ratio. We analyze the topology generated from Ripple network and find that averagely each user has 13 DT neighbors located in different direction. Considering the trade off between communication cost and success ratio, we choose 5 DT neighbors to probe in our design.
The pseudo-code of the forwarding algorithm at an intermediate node is shown in Algorithm~\ref{Algorithm_MDT} in Appendix~\ref{sec:appendix}.

From the observation of real PCNs, only a very small portion of the transactions are with big amounts. In Ripple, only 10\% of the payments are of balances more than \$1,740 USD, and the median balances of payment are \$4.8 USD \cite{wang2019flash}. We further introduce a special design to treat large transactions.
If a payment $t$ exceeds a threshold, the sender randomly divides the large payment into several micropayments $t_1, t_2, ... t_k$, and assigns each sub-payment a random unique index. The sender treats these sub-payments as unrelated and independent payments. 
If the number of sub-payments received by the receiver is less than the number informed by the sender, we assume that this payment $t$ is failed.
To better preserve the value privacy, small payments are also assigned a random index. Thus, a malicious node seeing a payment going through it cannot determine whether it is the total value or just part of the value of a sub-payment. Since the index is randomly distributed, the adversary cannot estimate the number of sub-payments or the total value of the payment from the index.

\vspace{-2ex}
\subsubsection{New node joins.}
A PCN network topology is dynamic, as new nodes may boots up while some existing nodes may get offline. So we first consider a join protocol to deal with nodes booting up.

When a new node $w$ boots up and wants to join the network, it first needs to discover its direct neighbors and assigns its coordinate.
Here, the direct neighbor can be any node that $w$ trusts and wants to build a channel to make frequent transactions. In order to get its coordinate, $w$ first needs to get the coordinates of anchors from the web servers. Then it sends hop-count queries to its direct neighbors and conclude its hop-count to each anchor. For example, assume the hop-count information to an anchor $w$ collects from its direct neighbors is $\{h_1, h_2,...\}$. $w$ can deduce that its hop-count to this anchor is $min\{h_1, h_2...\}+1$.
With these information, $w$ can determine its own coordinate locally.

Then it will send a join request to its neighbor who is a DT node, and try to find all of its DT neighbors. To begin its search, it must find at least one neighbor working correctly in the system, say $v$. 
Node $w$ includes its coordinate in the join request and sends it to $v$. Now $v$ can begin a greedy routing to get to node $c$ which is closest to $w$. By the property of DT, the closest node to $w$ in the Euclidean space must be a DT neighbor. So $c$ is definitely $w$'s DT neighbor. Then $c$ sends a $JOIN \_rep$ back to $w$ along the reverse path. After receiving the $JOIN \_rep$, $w$ begins an iterative search to find other DT neighbors. After finding the paths to these DT neighbors, $w$ locally store the paths as virtual links. $w$ also needs to locally maintain the direct neighbor set $C_u$ and DT neighbor set $N_u$. 
Since $w$ boots up to be their new neighbor, they also need to update their node sets $C_u$ and $N_u$ in the memory.

\vspace{-1ex}
\subsubsection{Node leaving and other changes}
We now consider a maintenance protocol to deal with the situation when a node gets offline, new channels emerge, or some channels shut down. This protocol is designed to fix the structure of MDT. The MDT graph is correct only if every node knows all of its DT neighbors. So in WebFlow, each node $u$ queries some of its neighbors to see if they know mutual neighbors that node $u$ does not know, and then sends neighbor-set requests to them. If $u$ discovers a new neighbor from neighbor-set replies, it will send a neighbor-set request to this new neighbor if they are vertexes in the same simplex in $DT(C_u)$. Every node runs this maintenance protocol locally, and every time when a node finishes running it, it will wait for a time period $T_m$ until running it again. This helps to keep the MDT correct and guarantees that the forwarding protocol works well.

\subsection{Limitations}
While MDT-based WebFlow routing protocol provides a high success ratio and short forwarding paths for payment with low memory cost in the payment channel network, it does not achieve all the privacy goals defined in Section~\ref{sec:overview}. 
Since we use Euclidean distance to choose the node closest to the receiver, the coordinate of the receiver is exposed to all the nodes along the path as well as their DT neighbors and direct neighbors. Once the adversary controls some of these nodes, it can speculate what and where the sender and receiver are with some probability, although it is still not sure about the payment value. 
Even if we apply onion routing in WebFlow as Lightning network did~\cite{poon2016bitcoin}, each intermediary still knows the coordinates of the immediately preceding and following nodes. Transactions may still be tracked by malicious intermediary.

%% file: 7-WF2.0.tex
\section{WebFlow-PE Design}
\label{sec:W2.0}

In this section, we present the routing protocol of WebFlow-PE, a privacy enhancement version in WebFlow. 
The design of WebFlow-PE is consistent with MDT-based WebFlow except forwarding part.
We first provide a high-level overview and intuition behind WebFlow-PE and then provide a detailed protocol description.

\subsection{WebFlow-PE Overview}
One problem of the MDT routing algorithm in the basic design of WebFlow in Section~\ref{sec:W1.0} is that the coordinate of the receiver is exposed to all the nodes along the path as well as their neighbors.
If there are malicious nodes standing along the path, they can somehow track the transaction and undermine privacy.



We extend the basic design of WebFlow to hide the coordinates of all the nodes along the path and enhance user anonymity based on an innovation called \textit{distributed Voronoi Diagram}. As introduced in Sec.~\ref{sec:W1.0}, the Voronoi Diagram is the dual graph of the DT. It is very easy for a node running WebFlow to know the boundaries of its Voronoi cell: simply computing the bisectors of the line segments to all its DT neighbors. If every node knows its Voronoi cell boundaries, a distributed Voronoi Diagram is maintained.


\begin{figure}[t]
	\centering
	\includegraphics[width=0.45\textwidth]{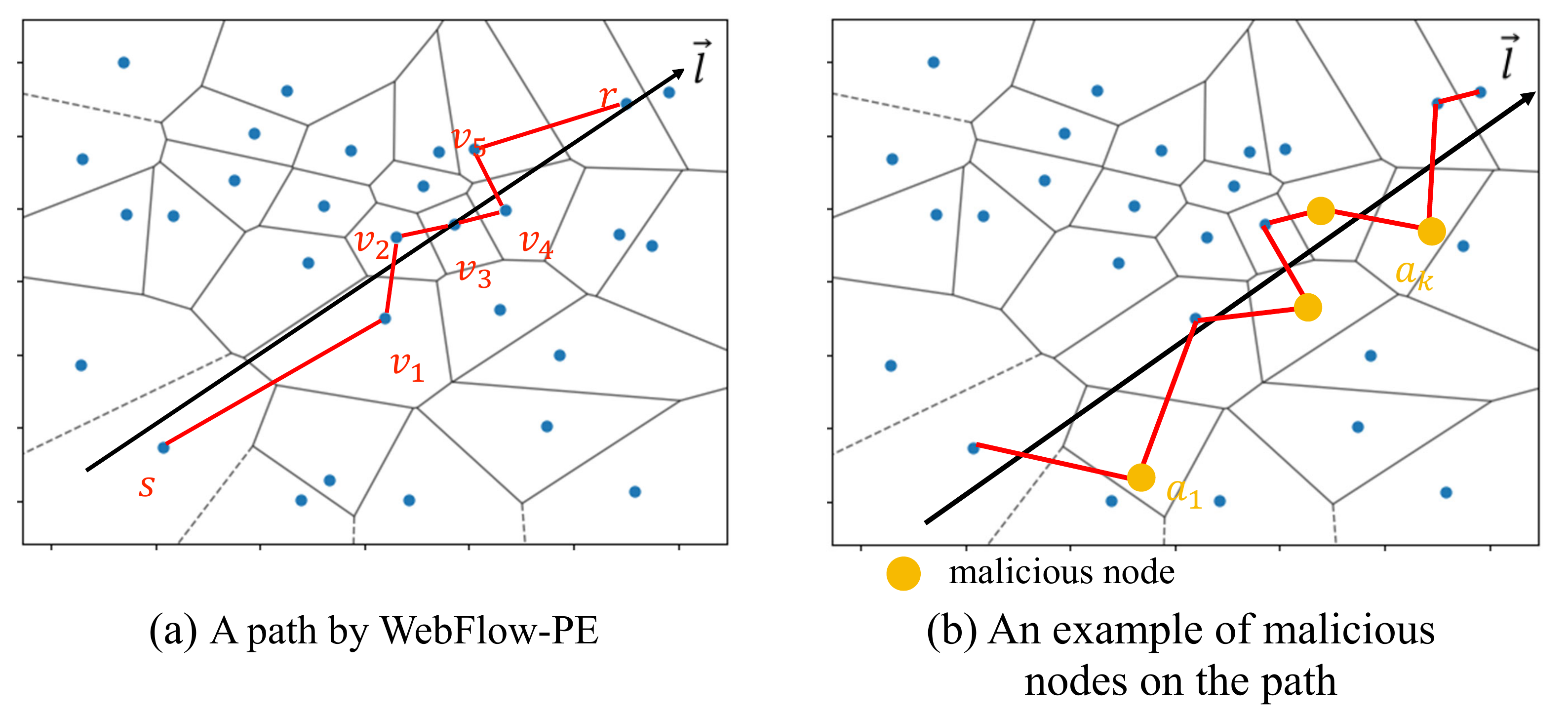}
	\vspace{-3ex}
	\caption{An example of WebFlow-PE on the distributed Voronoi diagram}
	\label{fig:WF-PE}
	\vspace{-2.5ex}
\end{figure}

\begin{figure*}[t]
	\centering
	\subfigure[\scriptsize{Channel Size Distribution of Ripple}]{
		\label{fig:ripple_channel}
		\includegraphics[width=0.23\textwidth]{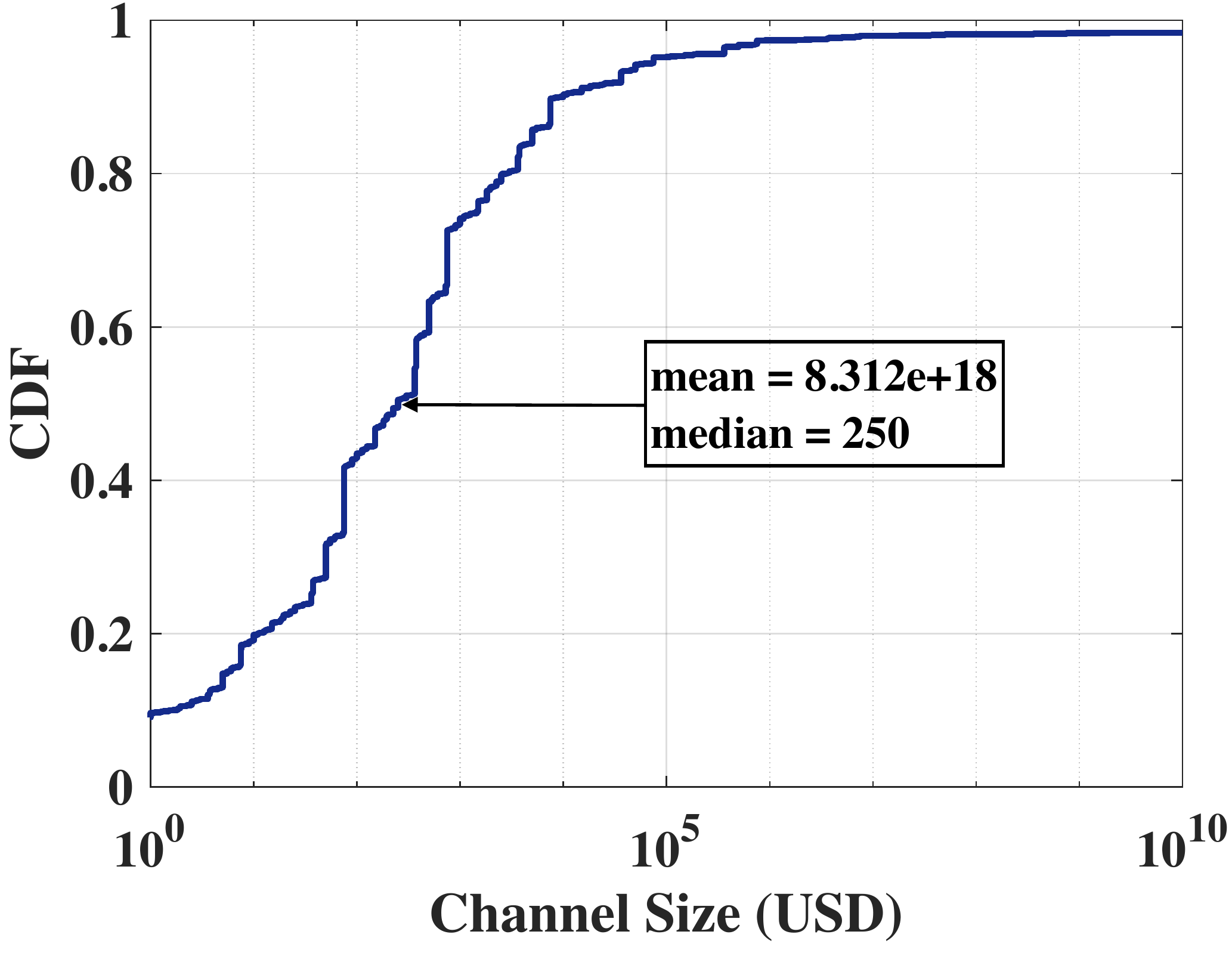}}
	\subfigure[\scriptsize{Transaction Size Distribution of Ripple}]{
		\label{fig:ripple_tx}
		\includegraphics[width=0.23\textwidth]{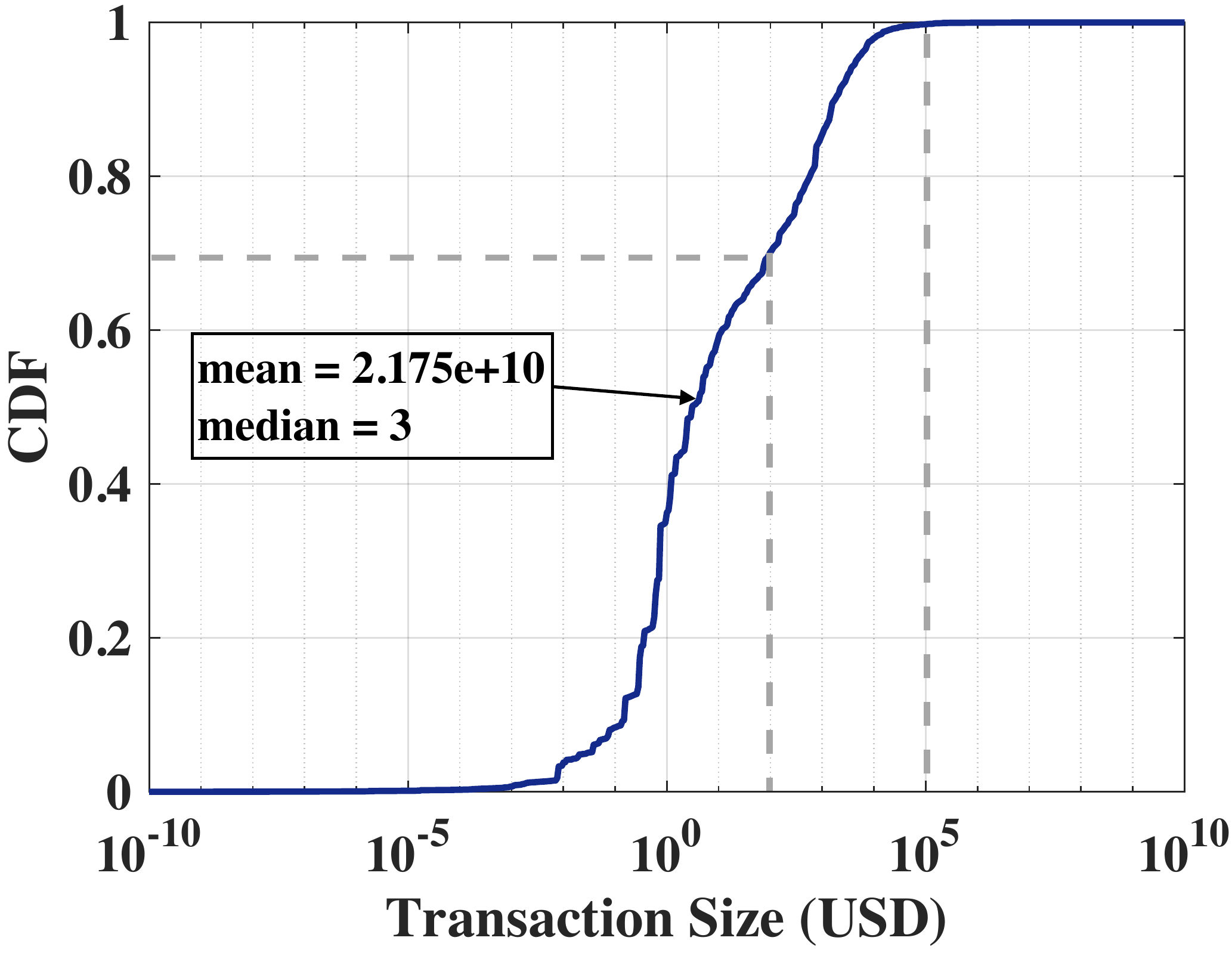}}
	\subfigure[\scriptsize{Channel Size Distribution of Lightning}]{
		\label{fig:lightning_channel}
		\includegraphics[width=0.23\textwidth]{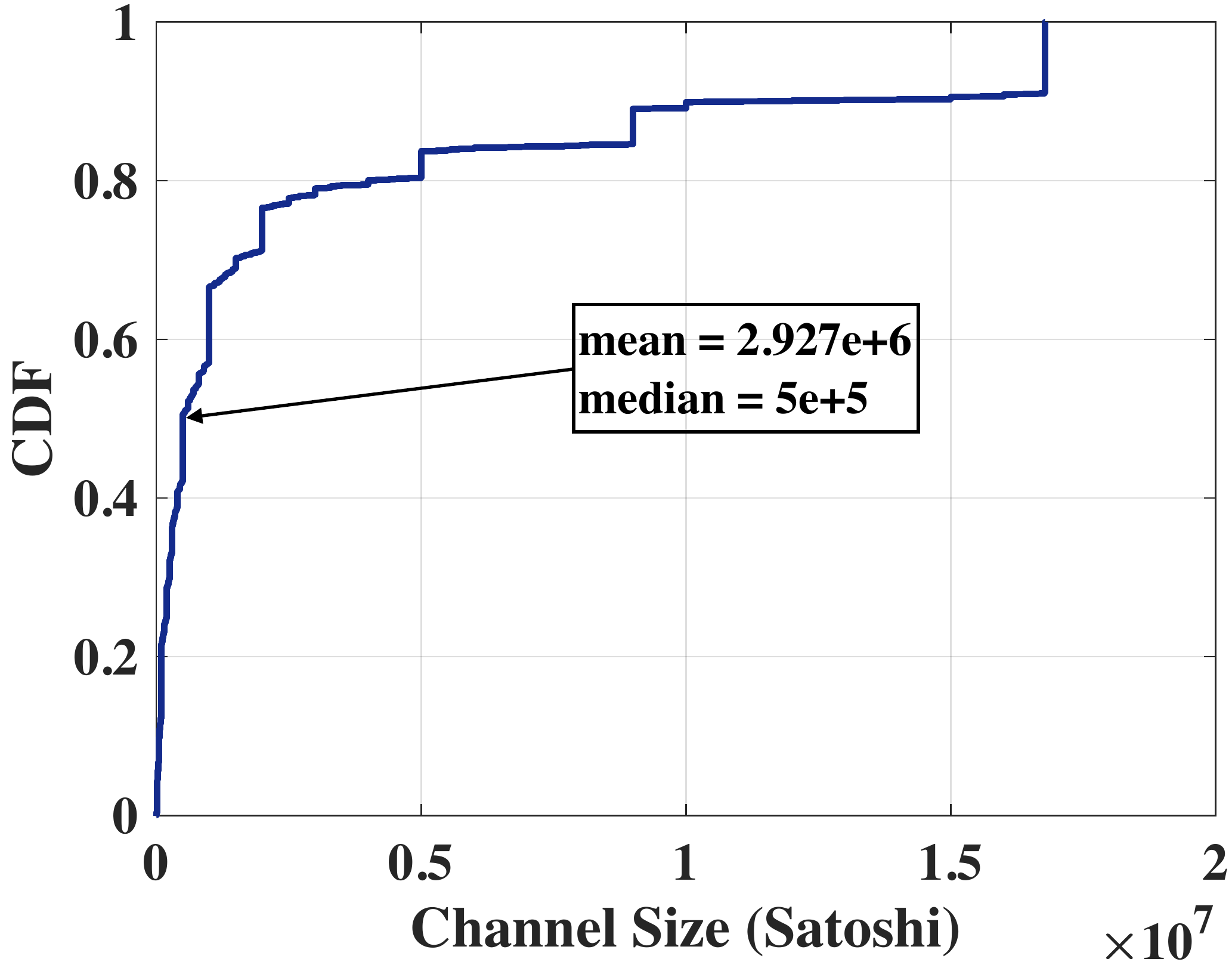}}
	\subfigure[\scriptsize{Transaction Size Distribution of Lightning}]{
		\label{fig:lightning_tx}
		\includegraphics[width=0.23\textwidth]{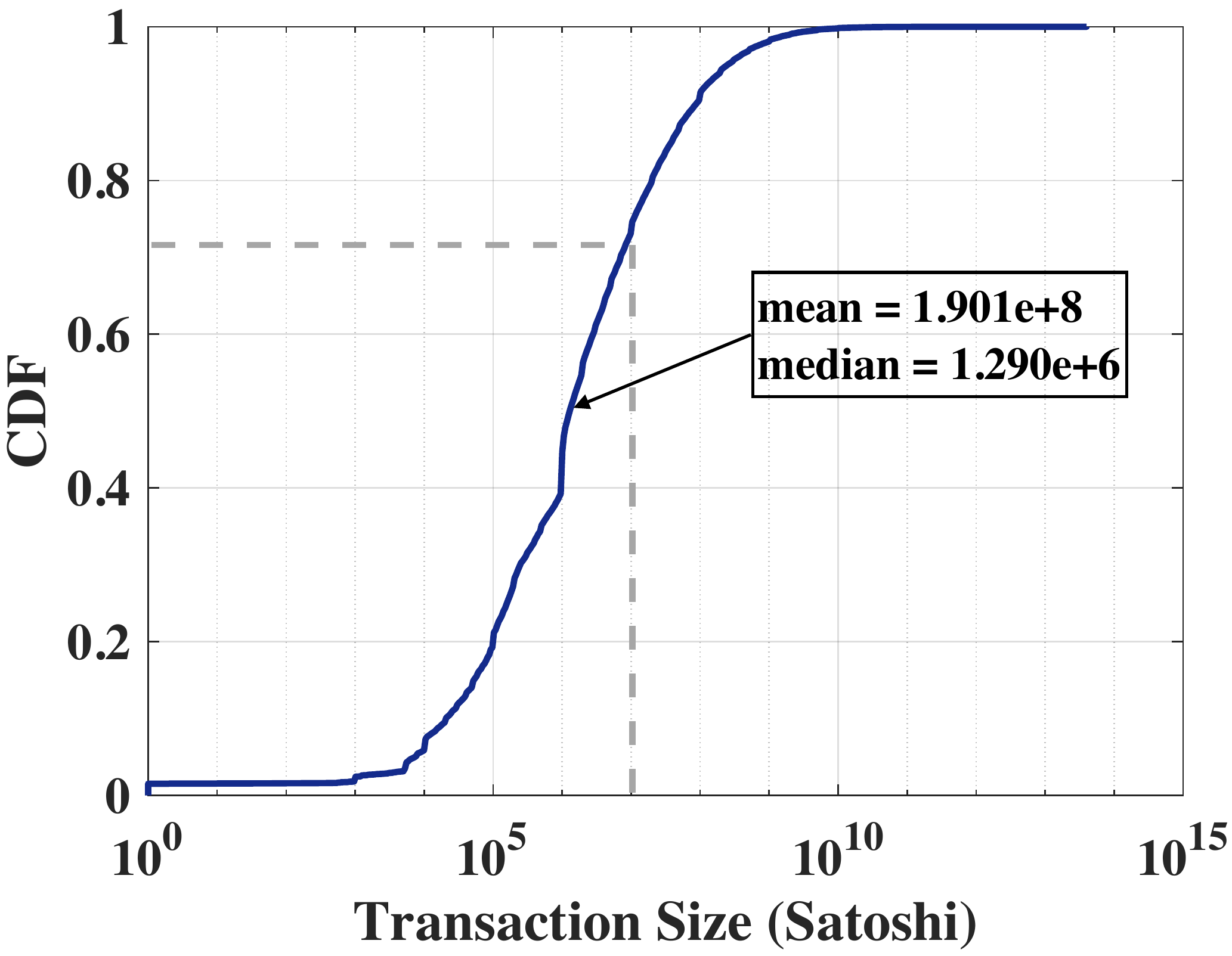}}
	\vspace{-3ex}
	\caption{\small{Transaction dataset and channel size distribution used for real-world evaluations.}}
	\label{fig:distribution}
	\vspace{-3ex}
\end{figure*}

Instead of using the destination coordinates, we propose to only use a line with a direction $\vec{l}$  as the routing target. 
The sender $s$ generates $\vec{l}$ by making it intersect an arbitrary point in the Voronoi cell of $s$ and another arbitrary point in the Voronoi cell of the receiver $r$, with a direction to the later as shown in Fig.~\ref{fig:WF-PE}(a).

The routing algorithm is that, at an intermediate node $v$, $v$ always sends the payment to its DT neighbor whose Voronoi cell is the next Voronoi cell intersecting with $\vec{l}$ along $\vec{l}$'s direction.
To prove that WebFlow-PE guarantees to find a path to $r$, we prove the following propositions. 

\vspace{-1ex}
\begin{proposition}
Suppose the Voronoi cell of $v$, $VC(v)$, intersects with $\vec{l}$.  The next Voronoi cell intersecting with $\vec{l}$ along $\vec{l}$'s direction is the Voronoi cell of  $v'$, denoted by $VC(v')$. Then $v$ and $v'$ are DT neighbors. 
\end{proposition}

\vspace{-1.5ex}
\begin{proof}
	Since $VC(v')$ is the Voronoi cell intersecting with $\vec{l}$ next to $VC(v)$,  $VC(v)$ and $VC(v')$ must share a
 Voronoi edge intersecting with $\vec{l}$. By definition, two nodes sharing a Voronoi edge are DT neighbors. 
\end{proof}

\vspace{-2ex}
\begin{proposition}
	$v$ knows a path to reach every DT neighbor of $v$. 
\end{proposition}

\vspace{-1ex}
This is very easy to prove: 1) if $v$ has a direct channel to its DT neihgbor $v'$, it can send the payment to $v'$ directly; otherwise 2) $v'$ is a multi-hop DT neighbor and $v$ can rely on MDT to get the path to $v'$. 

\vspace{-1ex}
\begin{proposition}
The routing of WebFlow-PE guarantees to reach the destination $r$.  
\end{proposition}

\vspace{-3ex}
\begin{proof}
Let the Voronoi cells intersecting with $\vec{l}$ are in a sequence $VC(s), VC(v_1), VC(v_2), ...,$ $VC(v_k), VC(r)...$. Since $\vec{l}$ intersects with $VC(s)$ and $VC(r)$ by definition, $VC(s)$ and $VC(r)$ must exist in the above sequence.
WebFlow-PE routing using $\vec{l}$ will visit $VC(s)$, $VC(v_1)$, $VC(v_2)$, ..., until reaching $VC(v_k)$ according to the routing algorithm. 
\end{proof}

\vspace{-1ex}
\textbf{How $r$ confirms that it is the receiver of the payment.} $s$ and $r$ need to first exchange a common secret for this transaction such as a transaction key $k$, using classic secure Internet communication such as TLS. In addition to $\vec{l}$, the sender will be a hash value $H(k)$ to the payment routing, where $H$ is a cryptography hash function. When $r$ receives a payment, it will compare  $H(k)$ to the hash of its transaction key and confirm that $\vec{l}$ intersects with $VC(s)$ and  $VC(r)$. After that, it can keep the payment and stop forwarding.



For each node along the path, it does not know the coordinates of the sender or the receiver. The only information it could obtain is $\vec{l}$. 
Each node can determine which DT neighbor is the next hop according to the direction function.
Besides, we also need to consider the capacity of the channel. 
After a node determines the next hop DT neighbor, it needs to check if the channel or multi-hop path has enough capacity to support the payment. If not, the payment fails.
We show the pseudo-code of the forwarding protocol of WebFlow-PE at each node $u$ along the path in Algorithm~\ref{Algorithm_voronoi} in  Appendix~\ref{sec:appendix}.

Different from MDT-based routing protocol, we first find a path for the payment and then probe the path to see if it has enough capacity to support the payment. Since the path is pre-determined, and it is static routing, it is more likely to fail than MDT-based routing protocol which is dynamic routing that combines the probing and path finding process at the same time. 
We mentioned that the direction function $\vec{l}$ can be the equation of any line intersecting the Voronoi regions of the sender and receiver. If the path formed by the first chosen $\vec{l}$ cannot support the payment, we may slightly adjust the direction function $\vec{l}$ to form another path. But the question of this naive thought is that it will not change the path a lot. And we cannot determine whether it will increase the maximum capacity of the path or decrease it, unfortunately. So it is not worth it to find other paths by generating some different direction function $\vec{l}$. It may not help a lot or even worsen the case while increasing the probing overhead. So we simply assume a payment failed if the path formed by the first chosen $\vec{l}$ does not have enough capacity to support the payment.

Note that the only difference between MDT-based WebFlow and WebFlow-PE is the forwarding protocol, and all the other designs keep the same. So the users of WebFlow can easily switch between these two routing protocols based on their demands, higher success ratio, or better privacy.

%% file: 8-Privacy.tex
\section{Privacy Analysis}
\label{sec:privacy}

\begin{figure*}[t]
	\centering
	\subfigure[Ripple]{
		\label{fig:state_ripple}
		\includegraphics[width=0.23\textwidth]{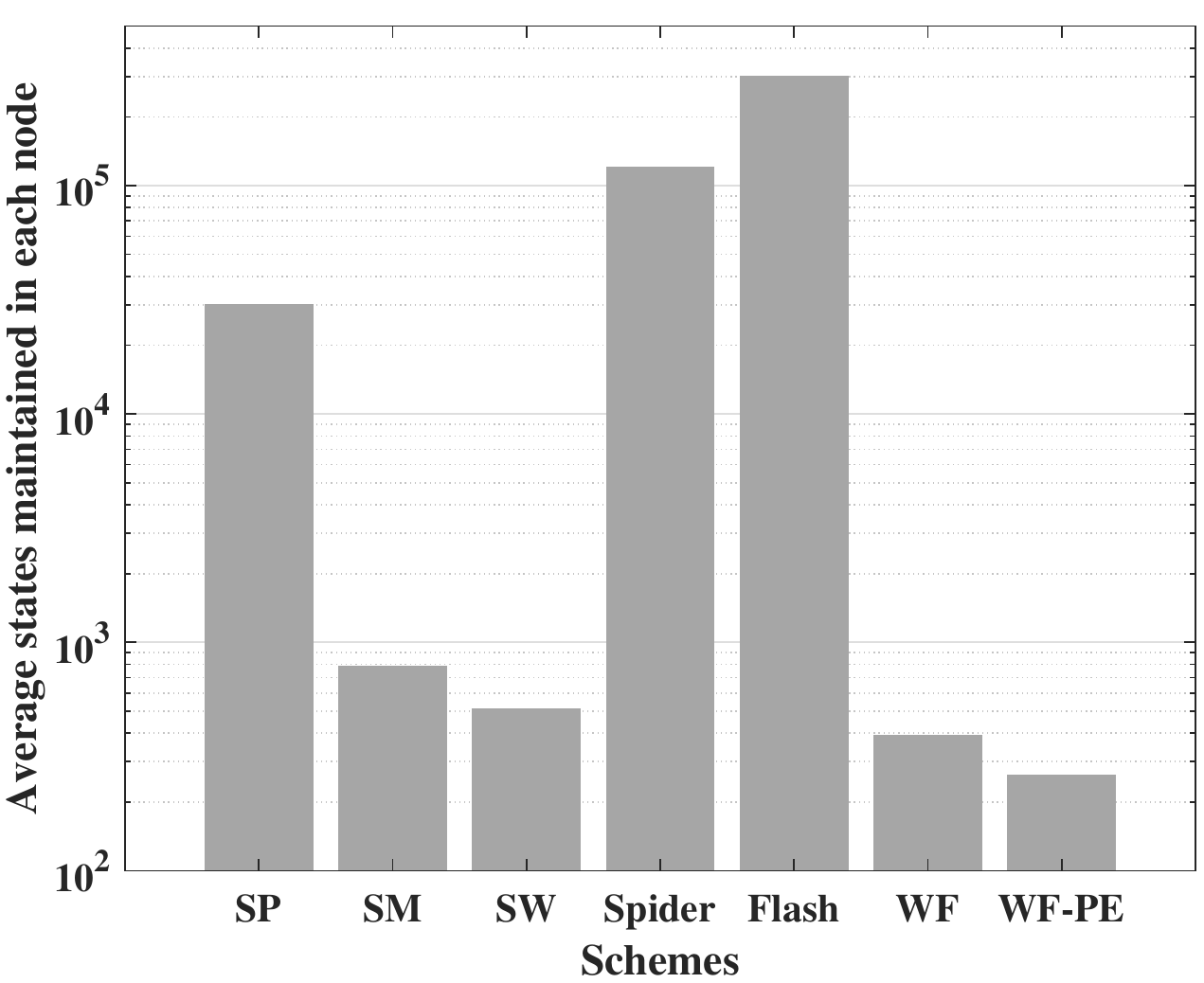}}
	\subfigure[Lightning]{
		\label{fig:state_lightning}
		\includegraphics[width=0.23\textwidth]{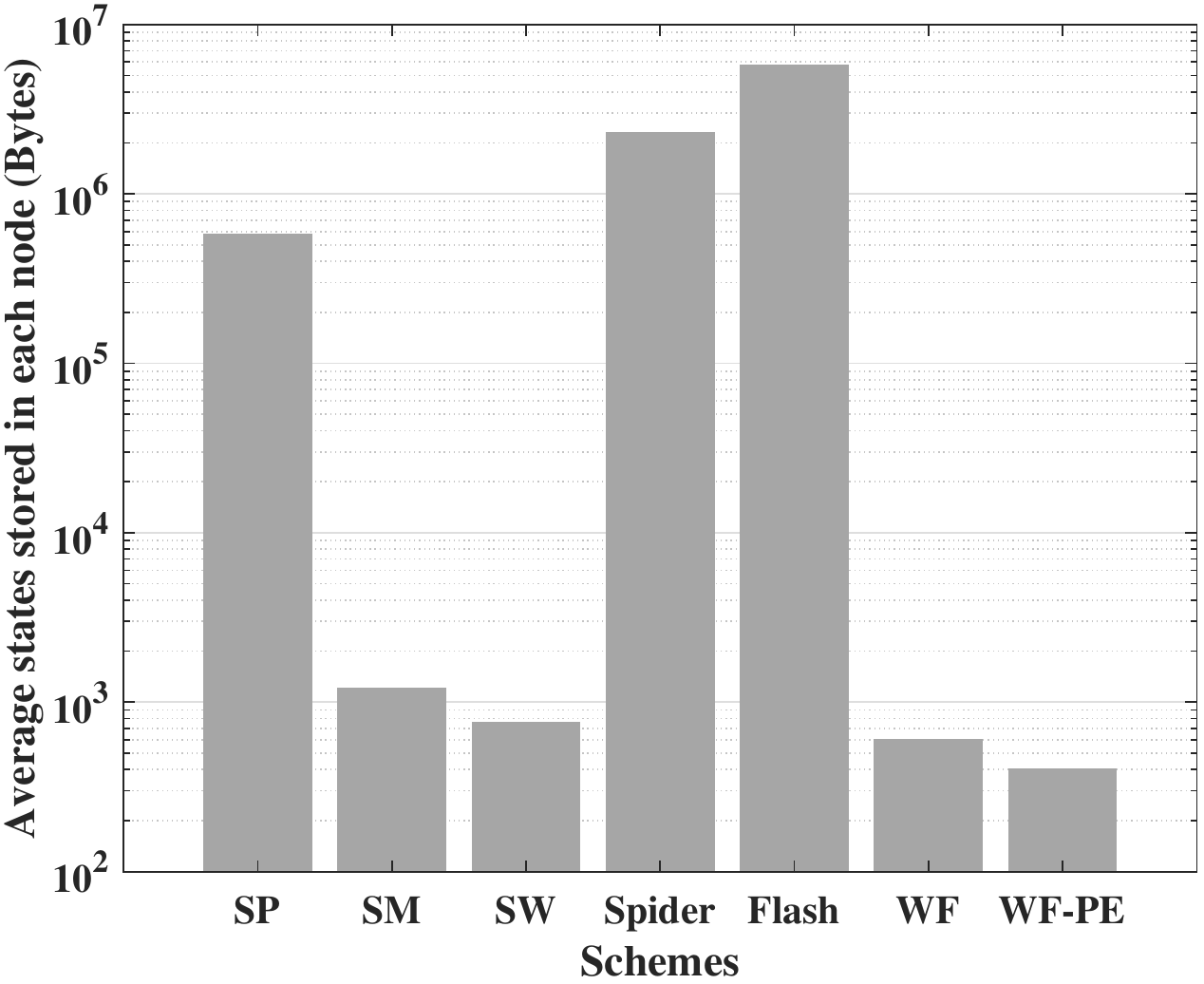}}
	\subfigure[Waxman]{
		\label{fig:state_waxman}
		\includegraphics[width=0.23\textwidth]{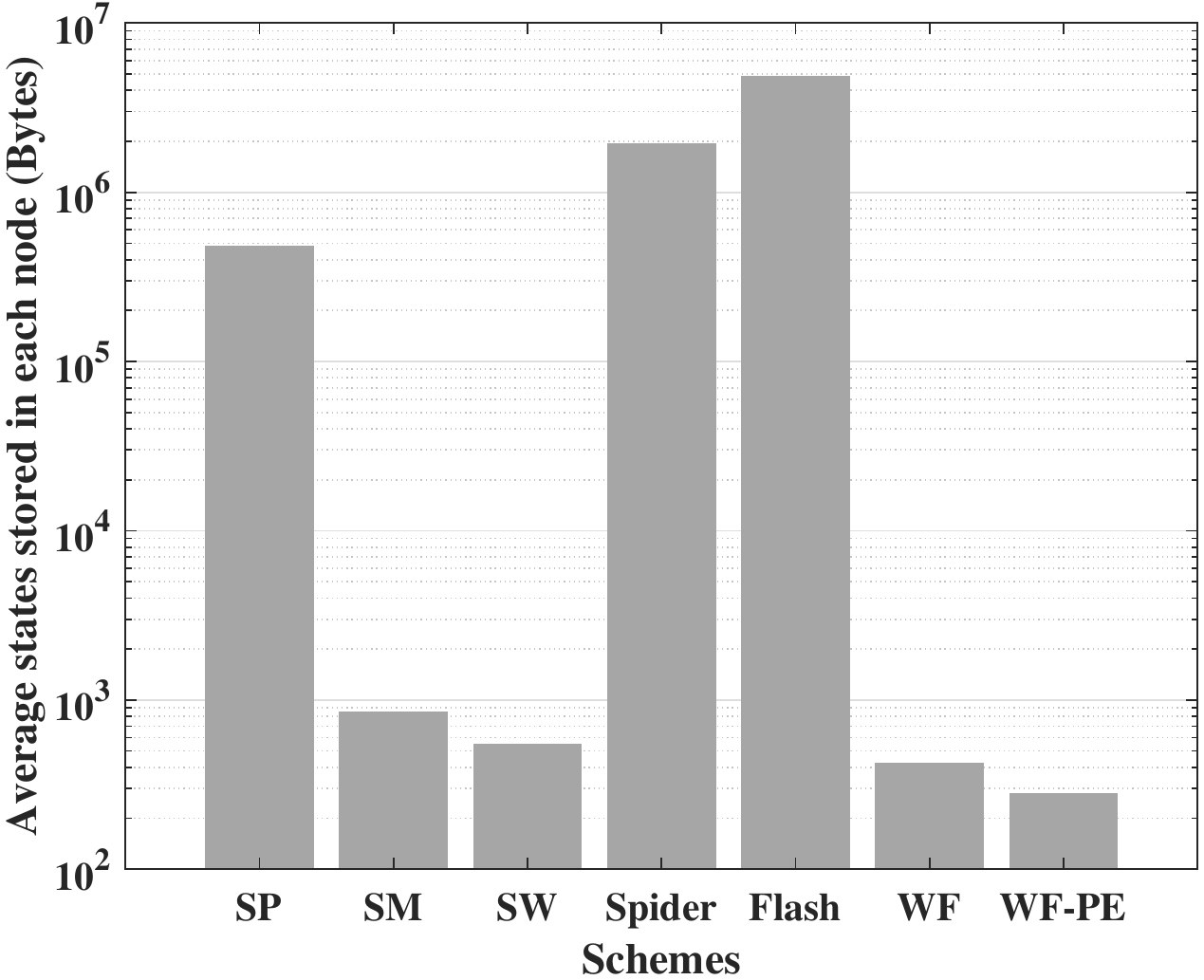}}
	\subfigure[Scale-free]{
		\label{fig:state_scale}
		\includegraphics[width=0.23\textwidth]{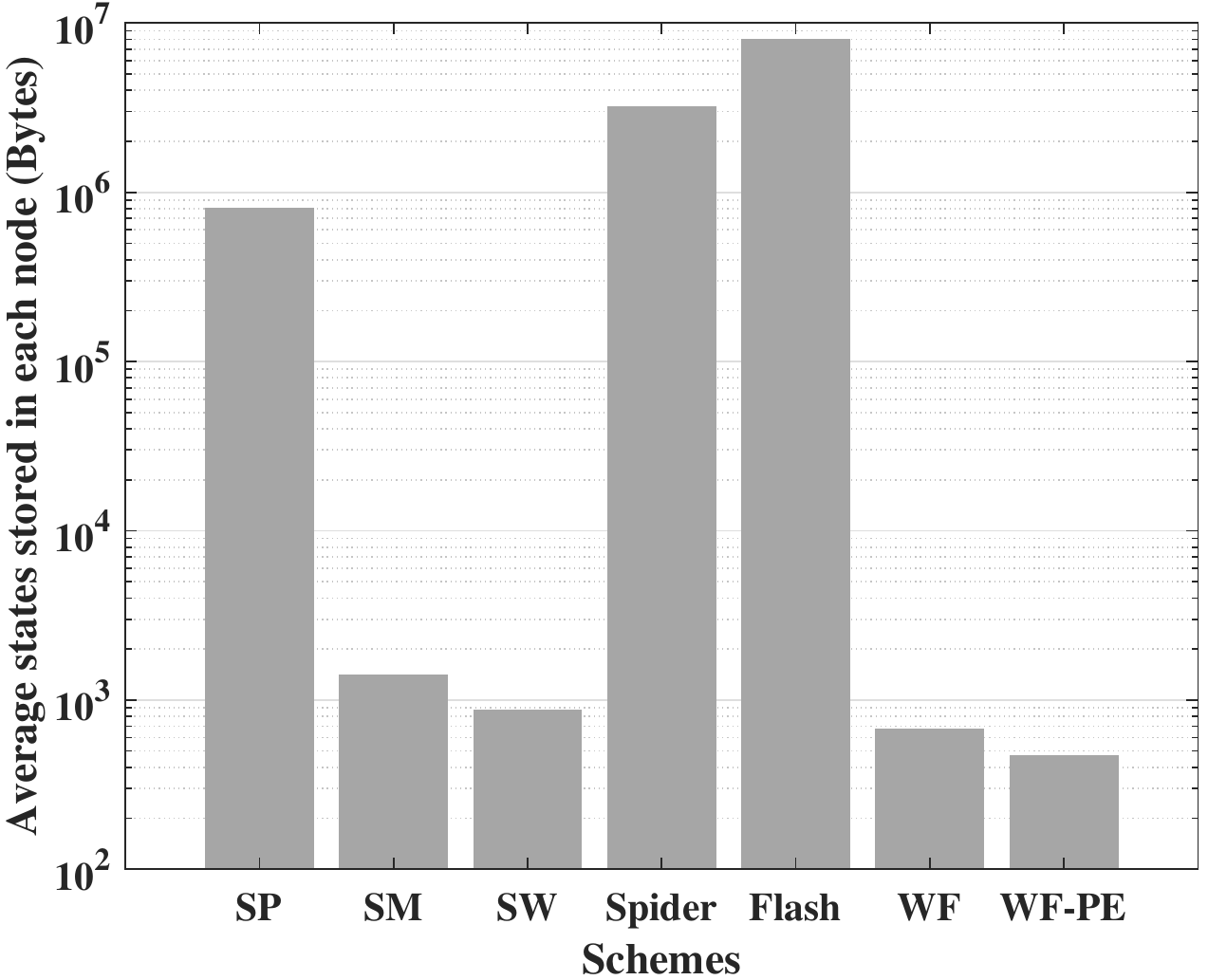}}
	\vspace{-3ex}
	\caption{Storage cost compared with benchmarks.}
	\label{fig:state}
	\vspace{-2ex}
\end{figure*}

\begin{figure*}[t]
	\centering
	\subfigure[Ripple]{
		\label{fig:probmsg_ripple}
		\includegraphics[width=0.23\textwidth]{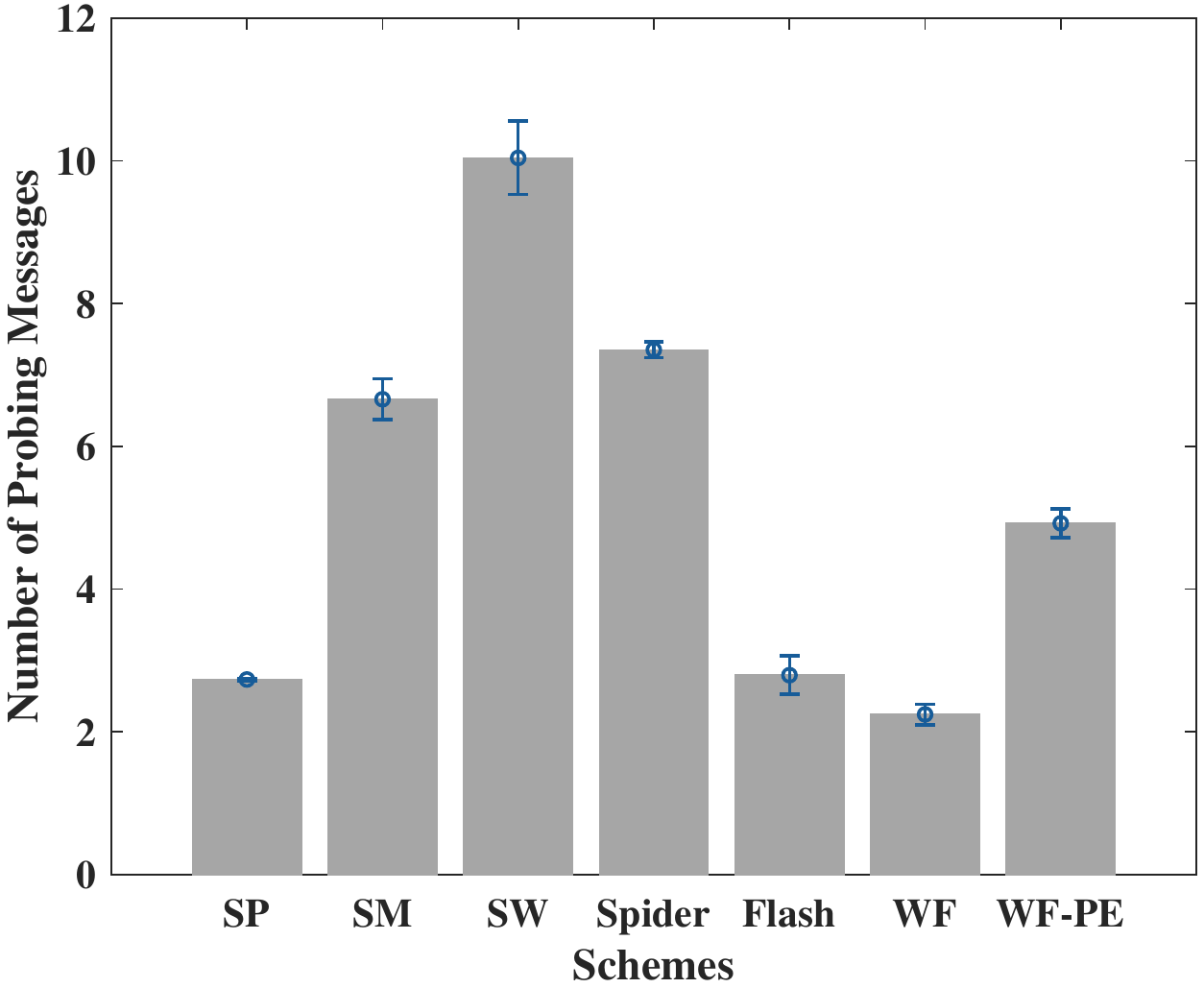}}
	\subfigure[Lightning]{
		\label{fig:probmsg_lightning}
		\includegraphics[width=0.23\textwidth]{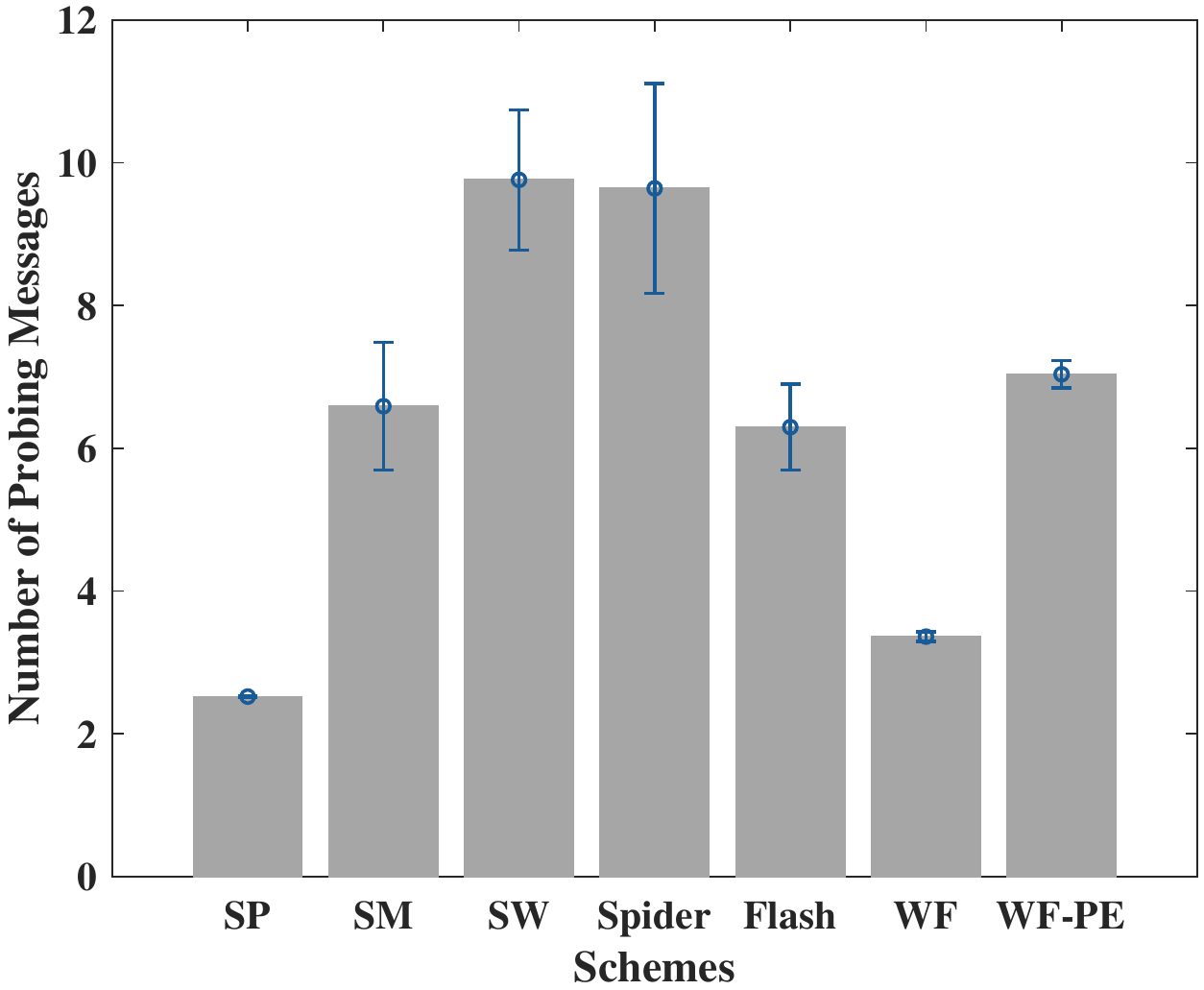}}
	\subfigure[Scale-free]{
		\label{fig:probmsg_scalefree}
		\includegraphics[width=0.23\textwidth]{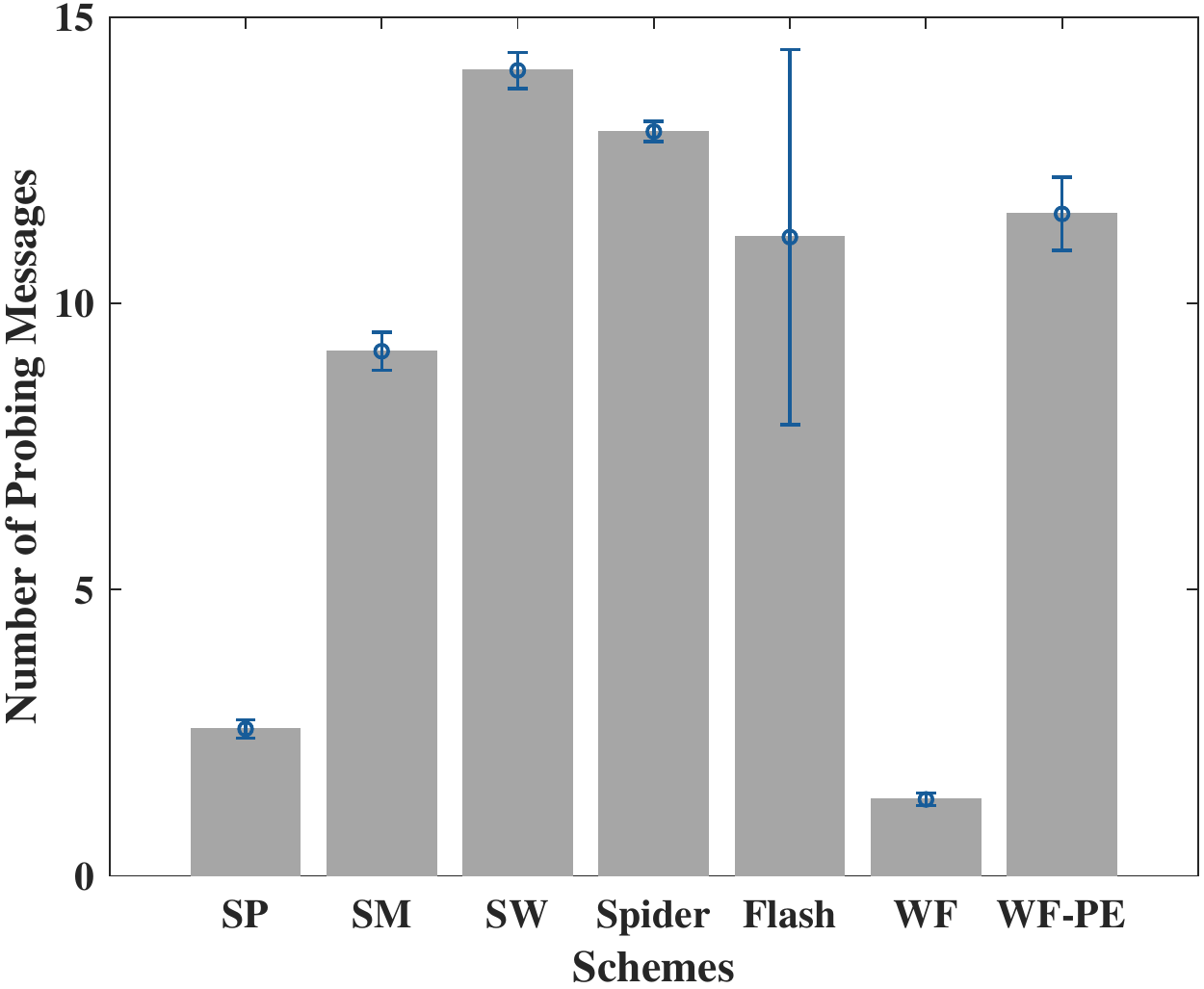}}
	\subfigure[Waxman]{
		\label{fig:probmsg_waxman}
		\includegraphics[width=0.23\textwidth]{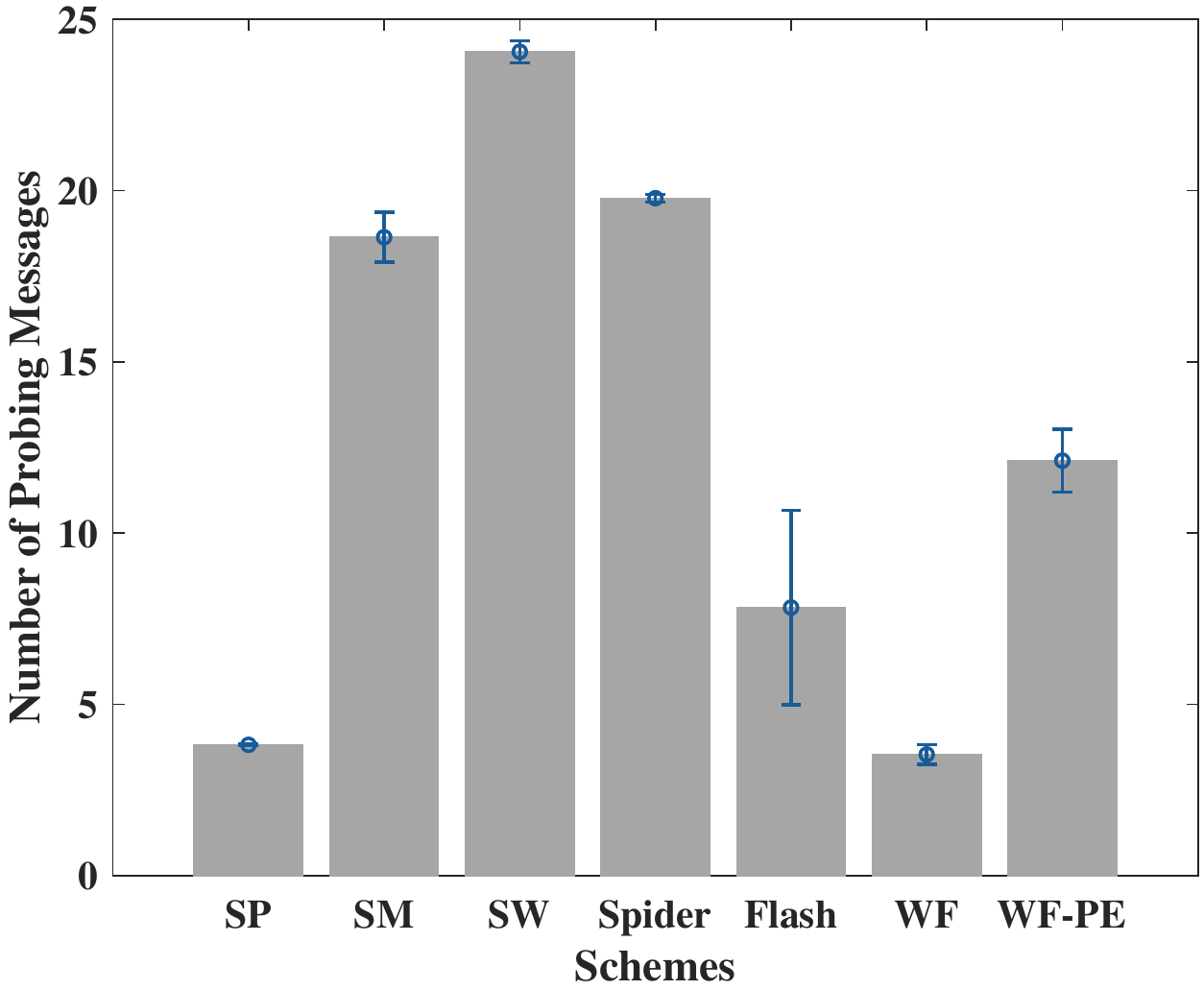}}
	 \vspace{-3ex}
	\caption{Communication cost compared with benchmarks.}
	\label{fig:probing}
	 	\vspace{-2ex}
\end{figure*}

We analyze the sender and recipient anonymity of our system in this section. Our anonymity measurement follows the anonymity definition by Pfitzmann et al.~\cite{zhuang2005cashmere}. We quantify WebFlow’s anonymity parameterized by:
\begin{itemize}
	\item $N$: network size,
	\item $f$ : fraction of malicious nodes in the network,
	\item $L$: number of relay groups in a path, the average path length,
	\item $\rho$: average node degree.
\end{itemize}
The anonymity measure uses the entropy metric defined in Sec.~\ref{sec:attmodel}, as $\frac{H(V)}{H_m(V)} = \frac{-\sum_{u \in V} p_u log_2(p_u)}{\log_2(|V|)}$.
We show why the entropy metric is better than the straightforward metric of the probability that the attacker knows the sender or receiver in  Appendix~\ref{sec:appendix}.

In MDT-based WebFlow, only the sender knows the source of payment. 
All nodes along the path as well as some of their neighbors only know the virtual coordinates of the recipient. So if there exist malicious nodes standing on the path, they can just guess some nearby nodes with a higher probability to be the recipient.
For those payments that do not go through the attacker $A$, all nodes look equally likely to be a sender to $A$. The probability of a path that avoids the attacker is:
\vspace{-1ex}
\begin{equation}
P = \frac{\tbinom{N*(1-f)}{L}}{\tbinom{N}{L}}
\end{equation}
For those payments that unfortunately go through the attacker, we can analyze the anonymity of the system using the entropy definition. 
The anonymity measure of the system is:
\begin{equation}
M = (1-P) \times \frac{-\sum_{u \in V} p_u log_2(p_u)}{\log_2(|V|)} + P \times 1
\vspace{-0.5ex}
\end{equation}
When there is an attacker on the path, if it's the last hop, it knows the destination for sure. Otherwise, it will guess that the probability of its nearby nodes to be the sender or recipient is $p_u = \frac{1}{L \times \rho}$. And for other nodes, they have the same probability 
$p_u = \frac{1-\frac{1}{L}}{N\cdot(1-f)-\rho}$ to be the sender or recipient.

In WebFlow-PE, same with MDT-based WebFlow, if a payment does not go through the attacker, anonymity can be guaranteed. What different is that, even if the adversary stands on the path of payment, it can only guess its next hop to be the recipient and its last hop to be sender with a probability $\frac{1}{L}$. 
If a path has more than one malicious node standing on it, these malicious nodes can collude to infer some information about this payment. For example, they know that all  nodes along the path between them cannot be the sender or the recipient. Since they know the average length of paths in the system according to historical data, they are more likely to infer the sender and recipient. Figure~\ref{fig:WF-PE}(b) shows an example. We assume that there are k malicious nodes on this path. For the node $a_1$, $a_k$ and one hop before/after $a_1$ and $a_k$ respectively, the probability of them to be the sender or recipient is: 
$p_u=\frac{1}{L-(k-2)}$. 
\vspace{0.1in}

For other neighbors of $a_1$, $a_k$, they cannot be the sender or recipient for certain. Since $a_1$, $a_k$ do not know who are the nodes or how many nodes between them along the path, all the other nodes except their neighbor looks equally to them. These nodes have the same probability to be the sender or recipient:
$p_u=(1-\frac{1}{L-(k-2)})\cdot \frac{1}{N-2\rho-2}$.
\vspace{0.1in}

Compared with both two versions of WebFlow, in landmark routing such as SilentWhispers, once a landmark is compromised, the system will have no privacy. 

We will show the simulation results on the comparison of privacy measures of different methods in the next section.

%% file: 9-evaluation.tex
\section{Performance Evaluation}
\label{sec:evaluation}


In this section, we evaluate the performance of WebFlow comparing to the existing off-chain routing algorithms, using \textbf{both simulation and prototype implementation}. The evaluation aims to answer the following research questions:

\begin{itemize}
	\item How does the WebFlow routing perform with regard to success ratio, success volume, and overhead under realistic PCN topologies and traces?
	\item How do link capacity and network load affect WebFlow’s performance?
	\item How do these results compare to the performance of other approaches?
\end{itemize}

\vspace{-0.1 in}
\subsection{Methodology}

We study WebFlow with two real-world PCN topologies: Ripple~\cite{armknecht2015ripple} and Lightning~\cite{poon2016bitcoin}, as well as synthesis topologies.
For real-world PCNs, We grab the data from January 2013 to November 2016, and
remove nodes with only a single neighbor and links with no funds from the topology. After processing, we get the network topology with 1,870 active nodes and 17,416 edges in our simulation. Similarly, we get the Lightning network topology with 2,511 nodes and 36,016 edges on one day of December 2018~\cite{wang2019flash}. Since the lightning network preserves the privacy of link balances, we cannot get the exact balance distribution on the link. However, we still can get the range of the link balance. Then, we evenly assign funds over both directions of a link. 

We generate payments by randomly sampling the Ripple transactions for the Ripple topology. Due to the lack of sender-receiver information in the trace of the Lightning network, we randomly sample the transaction volumes and sender-receiver pairs. The distribution of transaction sizes and channel size is shown in Figure~\ref{fig:distribution}. 70\% of the transactions in Ripple is less than 100$\$$, and 70\% of the transactions in Lightning is less than $10^7$ Satoshi. So we treat payments less than 100$\$$ and $10^7$ Satoshi as small payments in Ripple and Lightning respectively.
We assume that payments arrive at senders sequentially. Concurrent payments will be considered in the future work.

According to the observation of these two real-world topologies, we additionally build two sets of synthesis PCN topologies based on Waxman topology generation~\cite{waxman1988routing} and scale-free network model~\cite{scalefree}. The link balances are assigned similar to those of Ripple. The payments are also generated by mapping the Ripple transactions to the simulated topologies.

\begin{figure*}
	\centering
	\subfigure[Succ. ratio with all transactions]{
		\label{fig:dynamic_update_ratio_all_ripple}
		\includegraphics[width=0.23\textwidth]{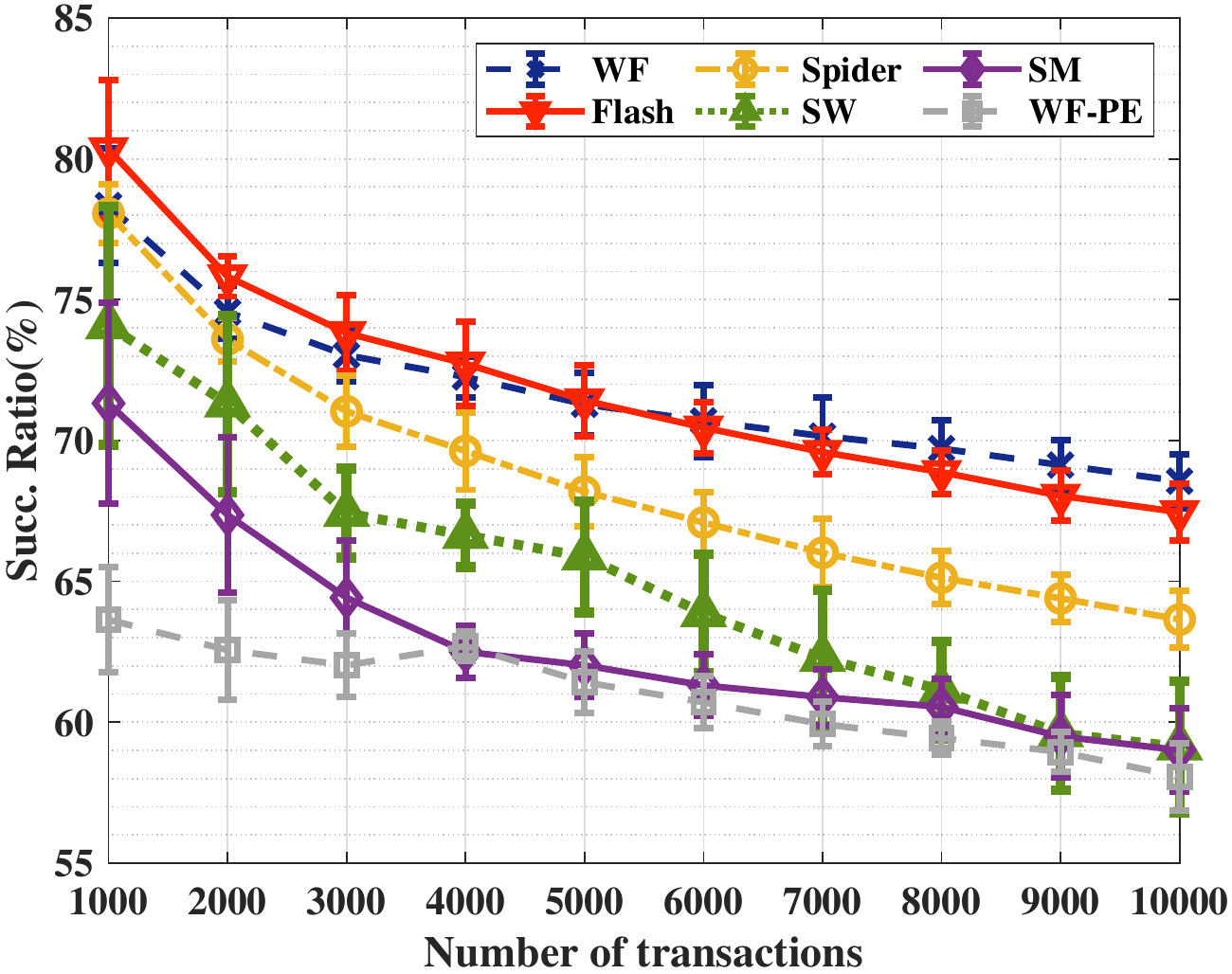}}
	\subfigure[Succ. volume with all transactions]{
		\label{fig:dynamic_update_volume_all_ripple}
		\includegraphics[width=0.22\textwidth]{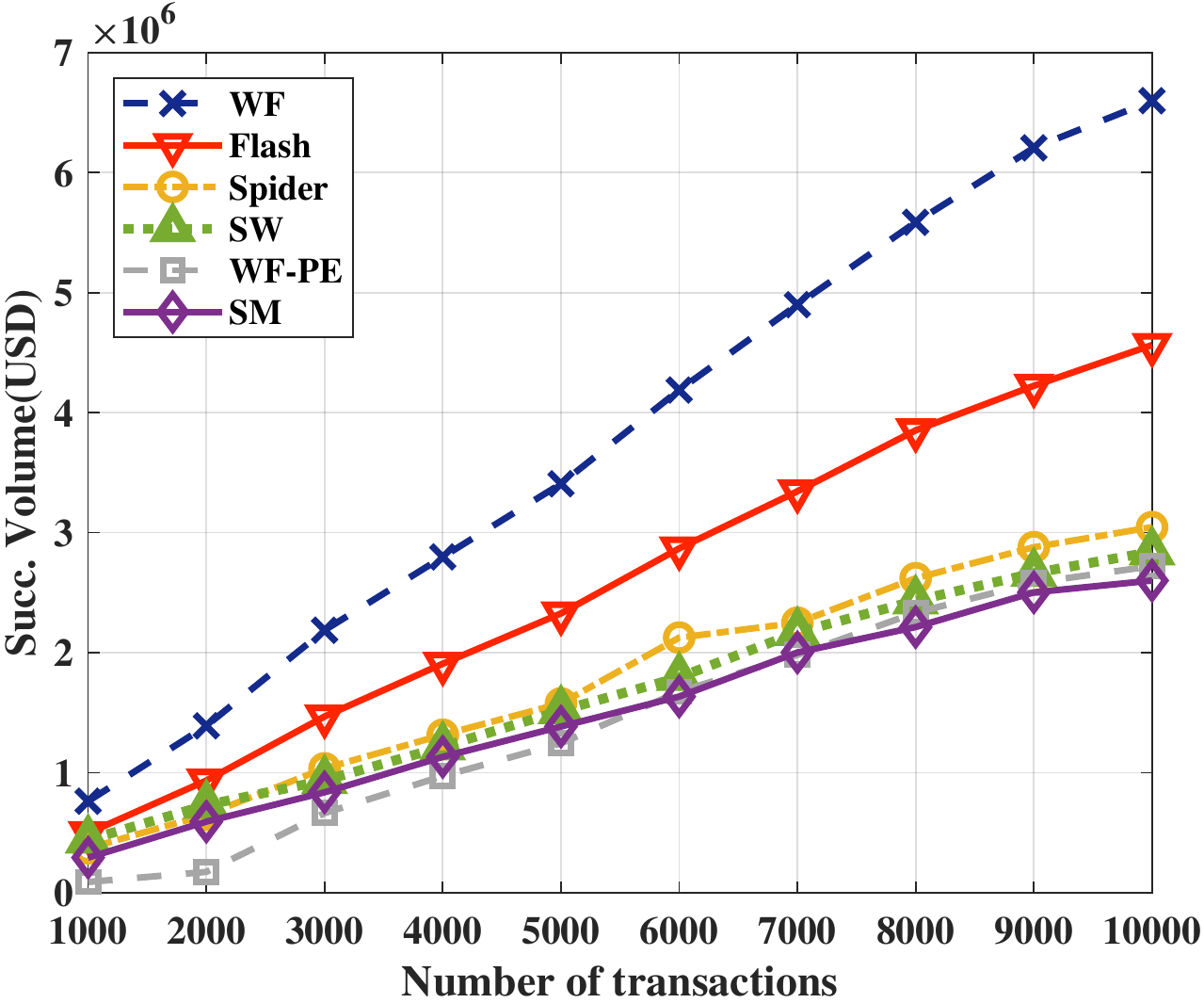}}
	\subfigure[Succ. ratio with small transactions]{
		\label{fig:dynamic_update_ratio_small_ripple}
		\includegraphics[width=0.22\textwidth]{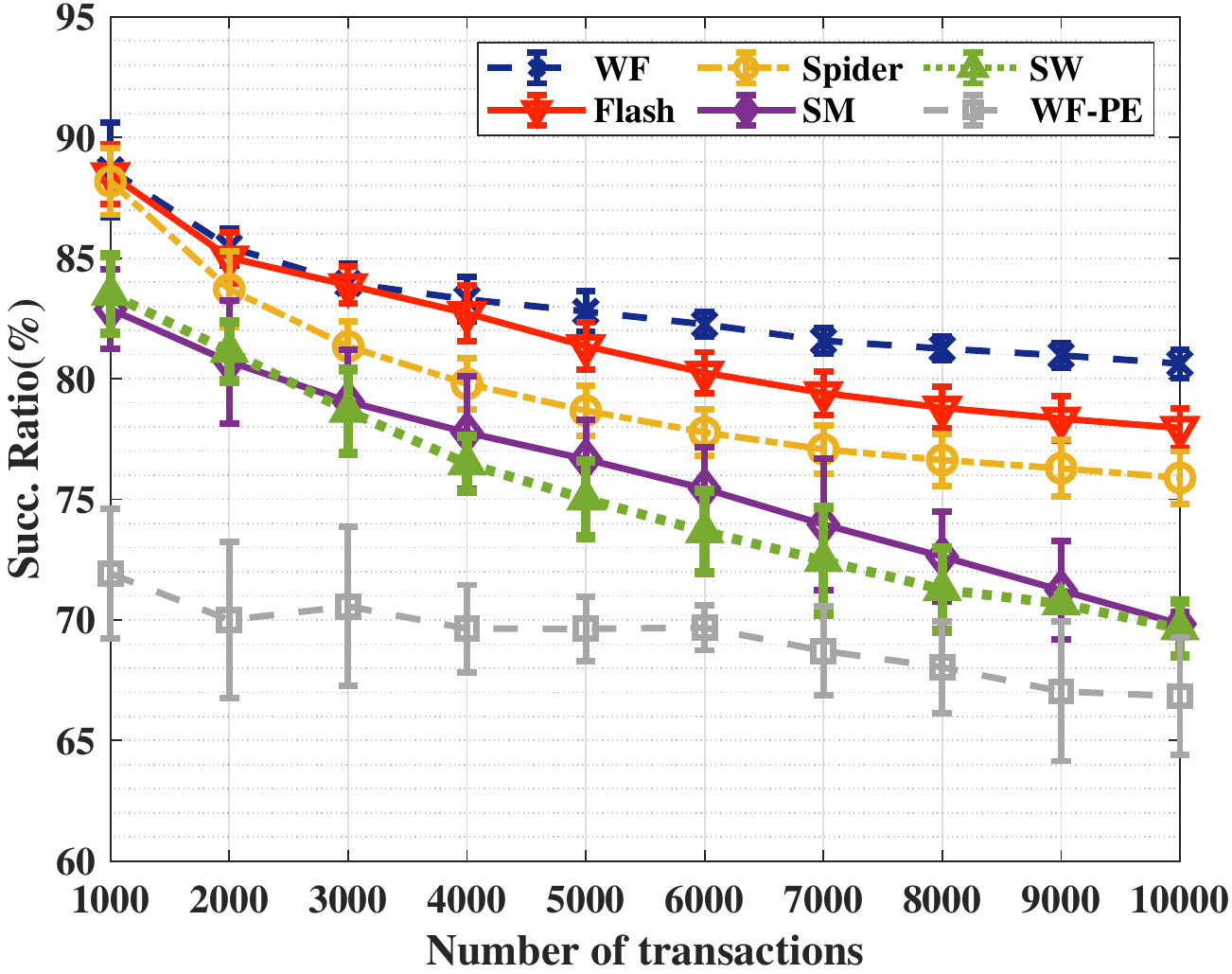}}	
	\subfigure[Succ. volume with small transactions]{
		\label{fig:dynamic_update_volume_small_ripple}
		\includegraphics[width=0.22\textwidth]{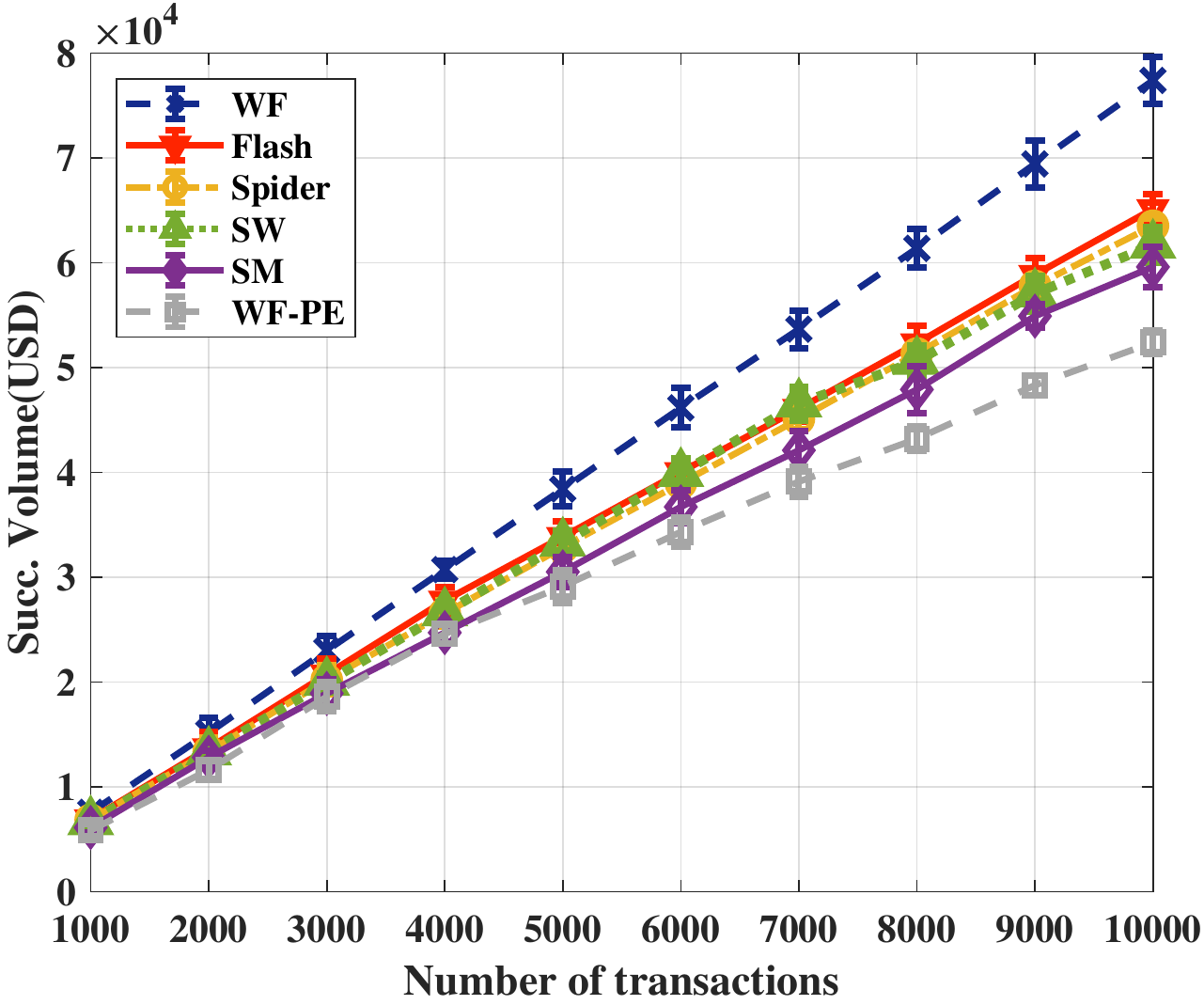}}
	\vspace{-2ex}
	\caption{Performance with varying transaction numbers in Ripple.}
	\label{fig:dynamic_ripple}
	\vspace{-3ex}
\end{figure*}

\begin{figure*}
	\centering
	\subfigure[Succ. ratio with all transactions]{
		\label{fig:dynamic_update_ratio_all_lightning}
		\includegraphics[width=0.23\textwidth]{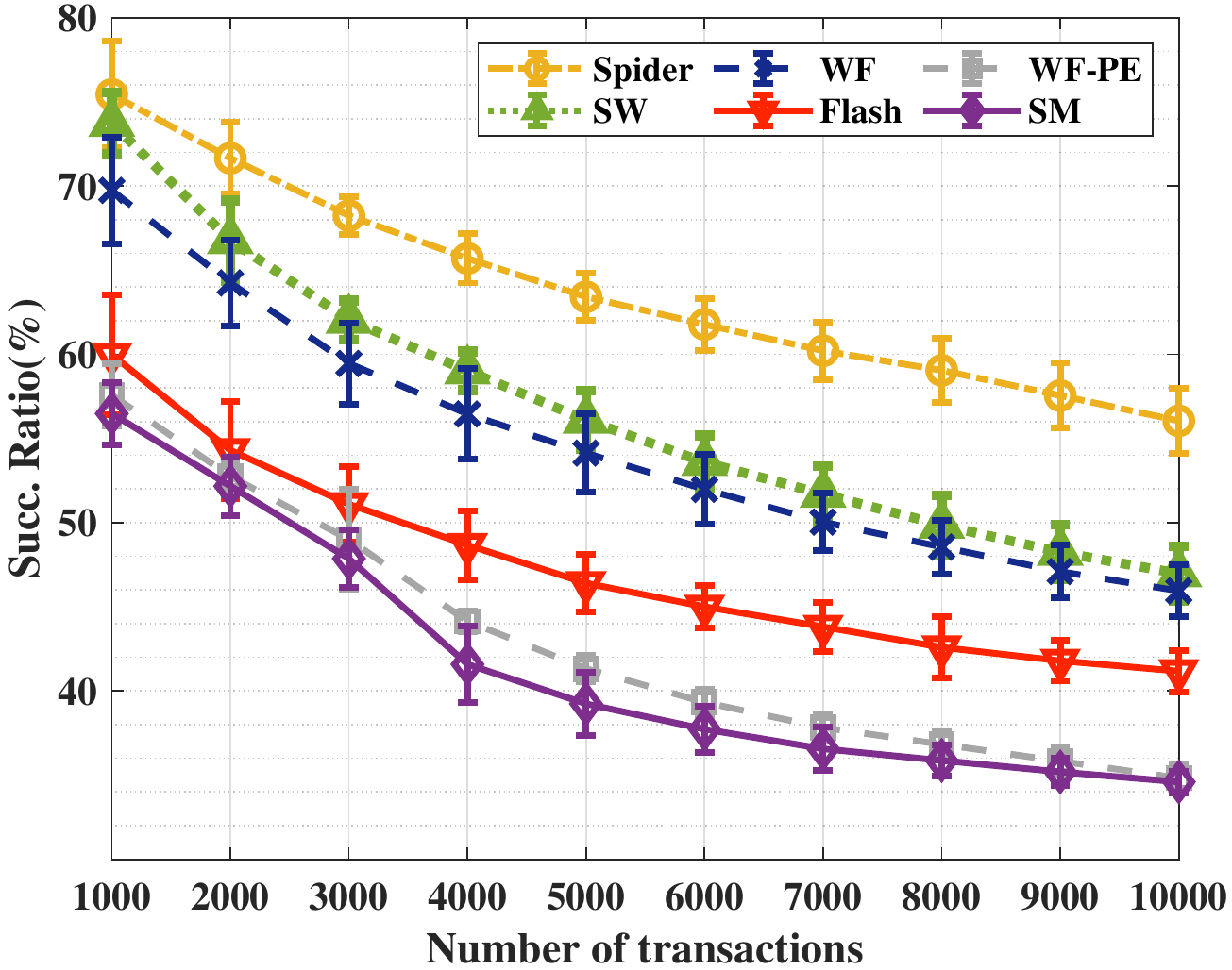}}
	\subfigure[Succ. volume with all transactions]{
		\label{fig:dynamic_update_volume_all_lightning}
		\includegraphics[width=0.22\textwidth]{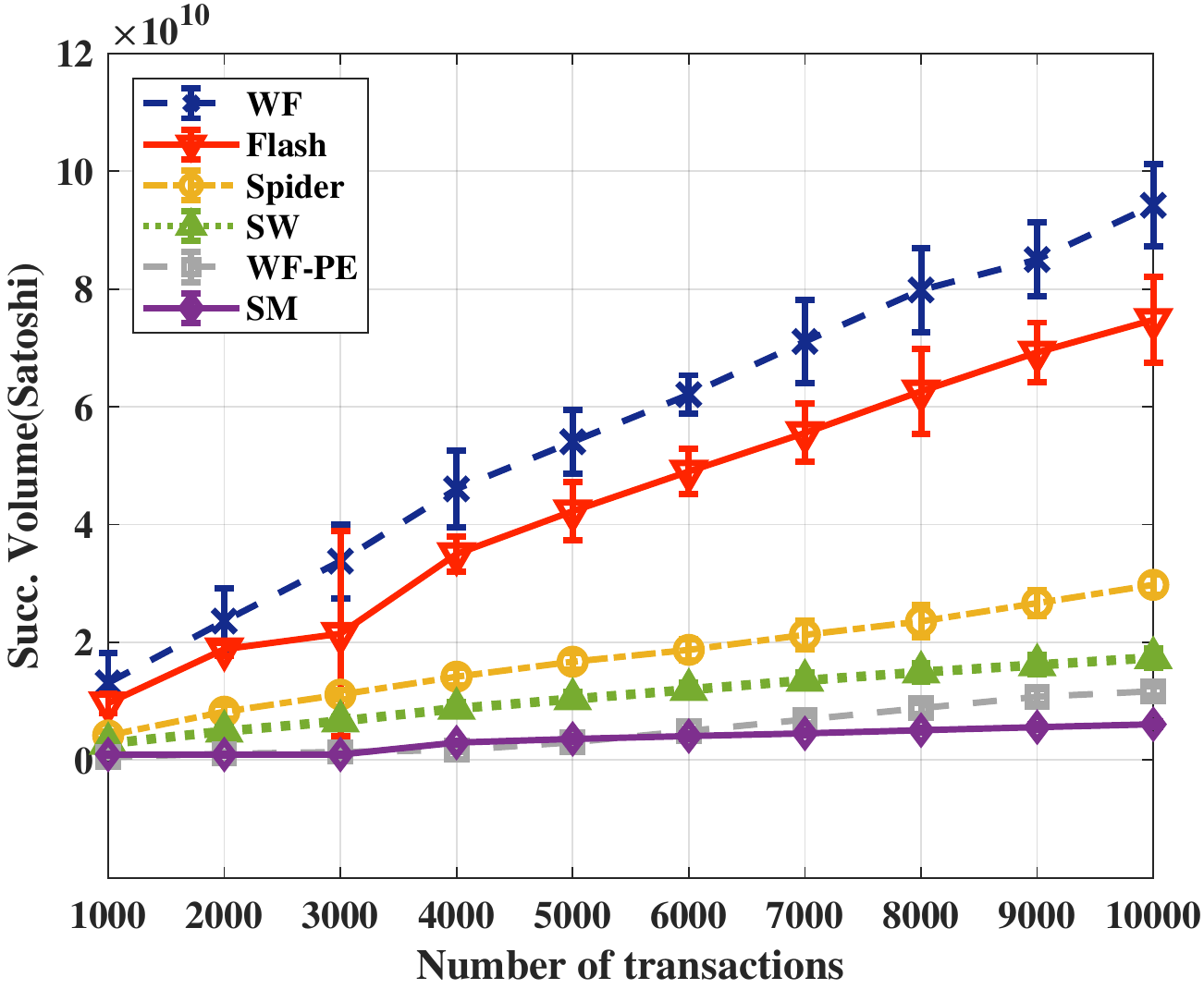}}
	\subfigure[Succ. ratio with small transactions]{
		\label{fig:dynamic_update_ratio_small_lightning}
		\includegraphics[width=0.22\textwidth]{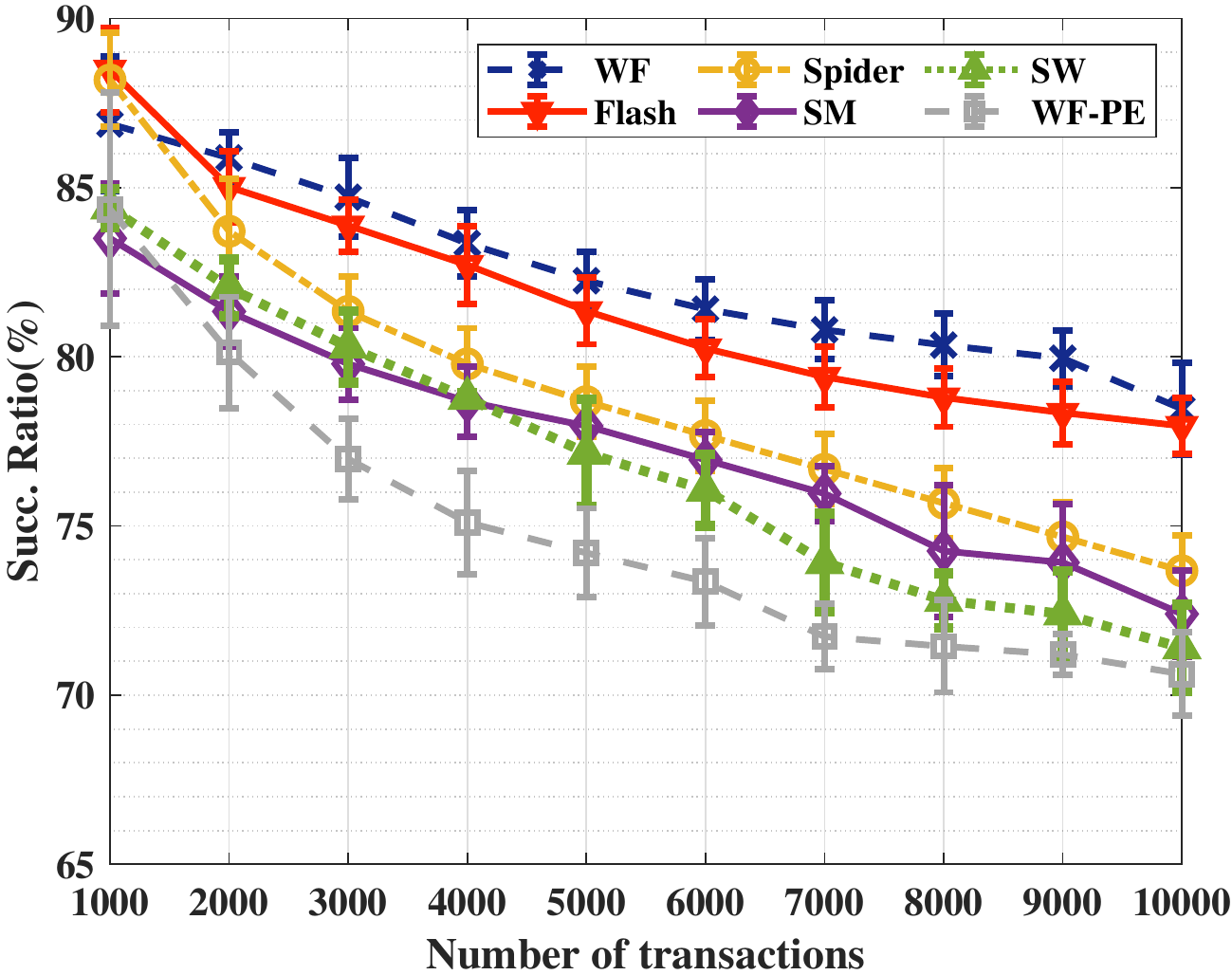}}	
	\subfigure[Succ. volume with small transactions]{
		\label{fig:dynamic_update_volume_small_lightning}
		\includegraphics[width=0.22\textwidth]{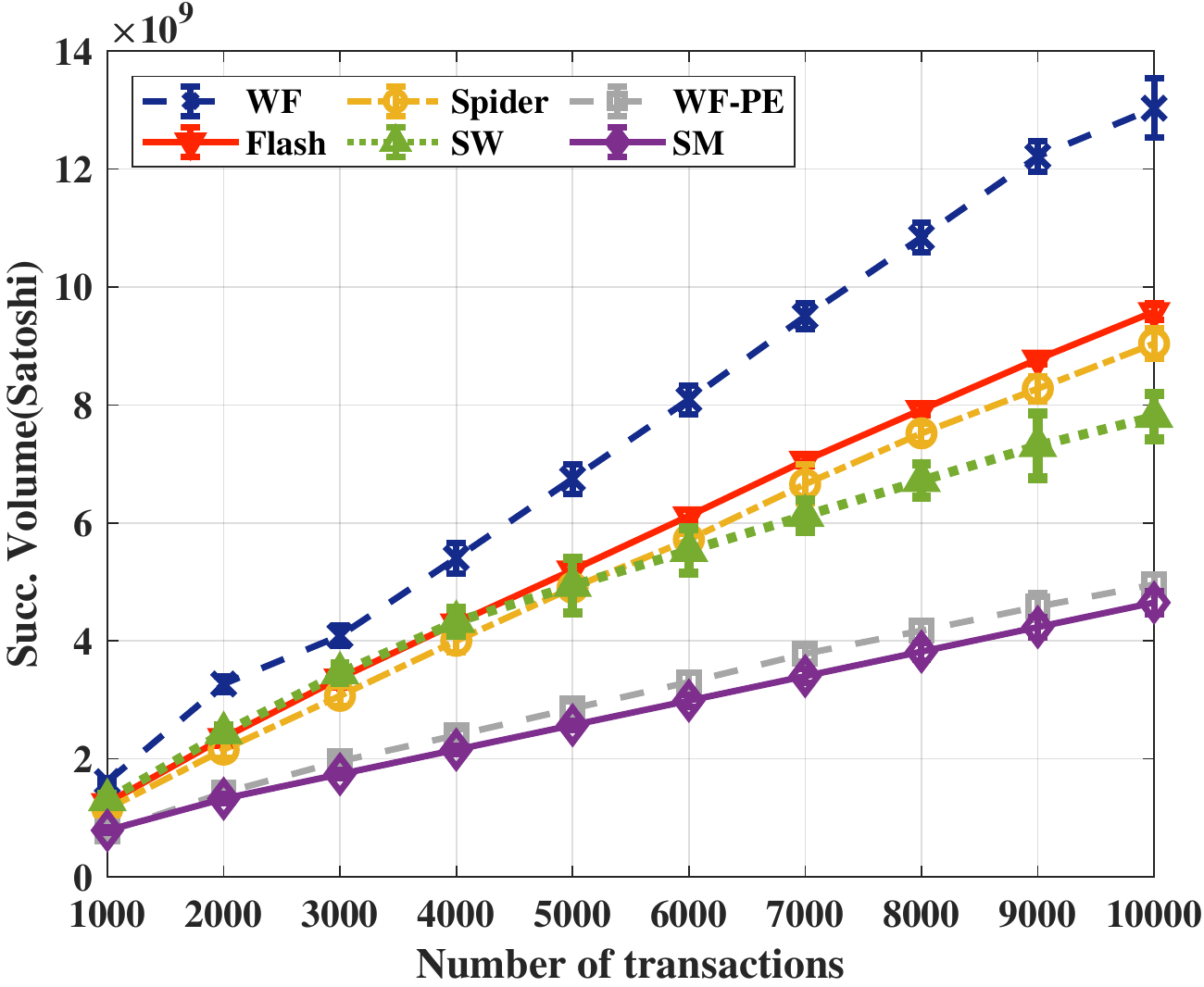}}
	\vspace{-3ex}
	\caption{Performance with varying transaction numbers in Lightning network.}
	\label{fig:dynamic_lightning}
	\vspace{-2ex}
\end{figure*}

\textbf{Benchmarks.} We compare both routing methods of WebFlow - MDT-based WebFlow (WF) and WebFlow-PE (WF-PE) - with the following off-chain routing algorithms.

\textit{SilentWhispers (SW)}~\cite{malavolta2017silentwhispers}:
A landmark-centered routing algorithm in which nodes always send funds to landmarks, and rely on landmarks to find the path. We set the number of landmarks to 3.

\textit{SpeedyMurmurs (SM)}~\cite{roos2017settling}:
An embedding-based routing algorithm that relies on assigning coordinates to nodes to find shorter paths with reduced overhead. We set the number of landmarks to 3 as ~\cite{roos2017settling} suggests.

\textit{Spider}~\cite{sivaraman2018routing}:
The off-chain routing algorithm which considers the dynamics of link balance. It balances paths by using those with maximum available capacity, following a waterfilling heuristic. It uses 4 edge-disjoint paths for each payment.

\textit{Flash}~\cite{wang2019flash}:
The off-chain routing algorithm which differentiates the treatment of elephant payments from mice payments. We set the number of preserved paths for each receiver in mice payment routing to 4, and the number of preserved paths for elephant routing to 20. The elephant-mice threshold is set such that 90 $\%$ of payments are mice.

\textit{Shortest-Path (SP):}
It serves as the baseline. SP uses the path with the fewest hops between the sender and receiver to route a payment.

\textbf{Metrics.}
We use communication and storage costs as the primary metrics for scalability. Similar to prior work~\cite{roos2017settling,wang2019flash}, we also use success ratio and success volume as evaluation metrics for resource utilization. Besides, we evaluate the anonymity of the system. 
The success ratio is defined as the percentage of successful payments whose demands are met overall generated payments. The success volume describes the total size of all successful payments, which is the throughput of the network. 
Before sending payments, nodes need to probe the usable capacity of the candidate paths. The number of probe messages describes the communication cost, which is the probing overhead. 
The storage cost is computed by the average number of distinct neighbors a node needs to know (and store) to perform routing, including the coordinates of its neighbors and the information of its related links.
The definition of anonymity of the system is defined in Section~\ref{sec:privacy}.
We report the average results over 5 runs.

\vspace{-1ex}
\subsection{Overall Performance and Overhead}
We now evaluate the performance of WebFlow with different settings of PCNs.

\begin{figure*}[t]
	\centering
	\begin{minipage}[t]{0.24\textwidth}
		\centering
		\includegraphics[width=0.98\textwidth]{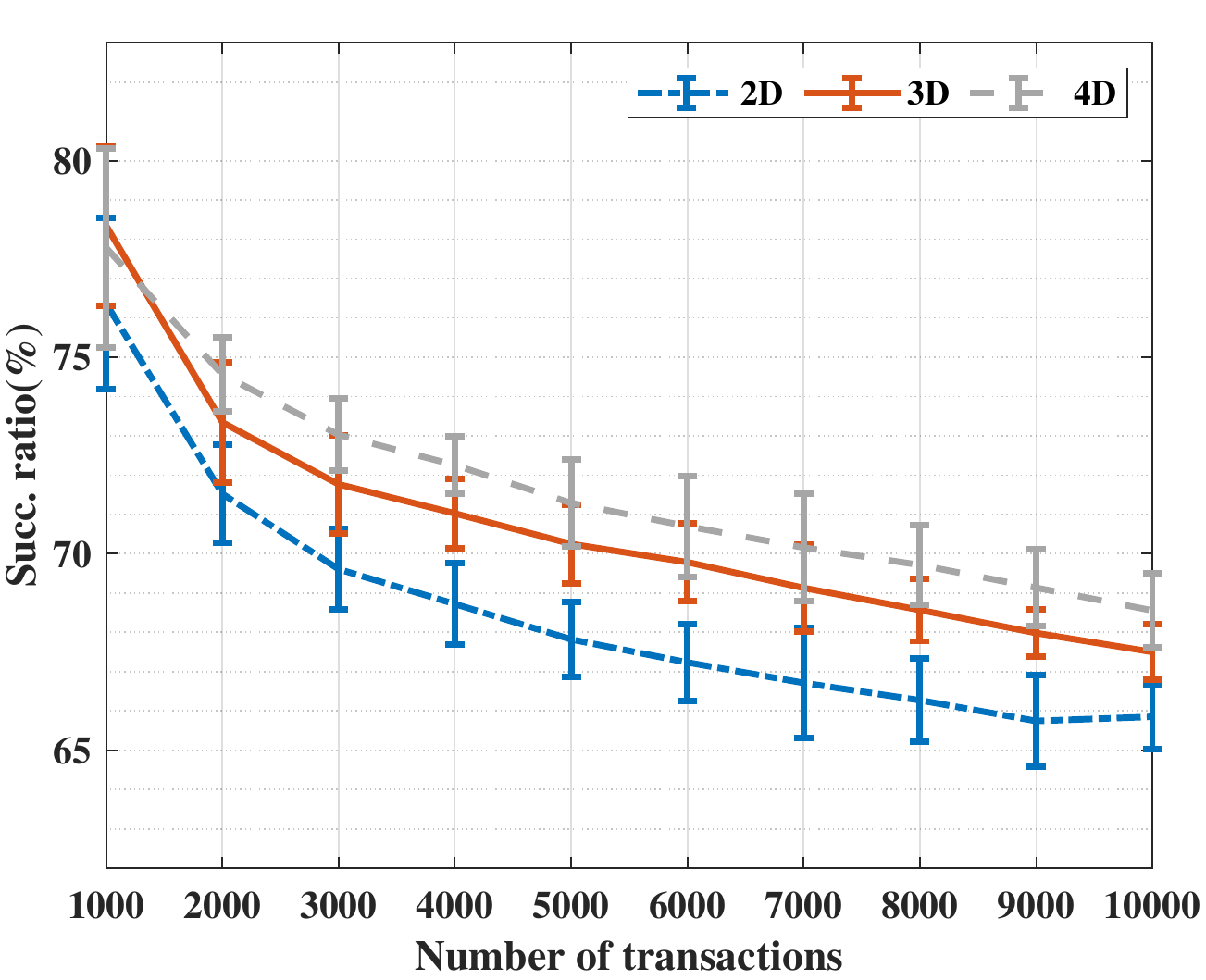}
		\vspace{-3ex}
		\caption{Routing succ. ratio in Ripple}
		\label{fig:dimension_ripple}
	\end{minipage}
	\begin{minipage}[t]{0.24\textwidth}
		\centering
		\includegraphics[width=0.98\textwidth]{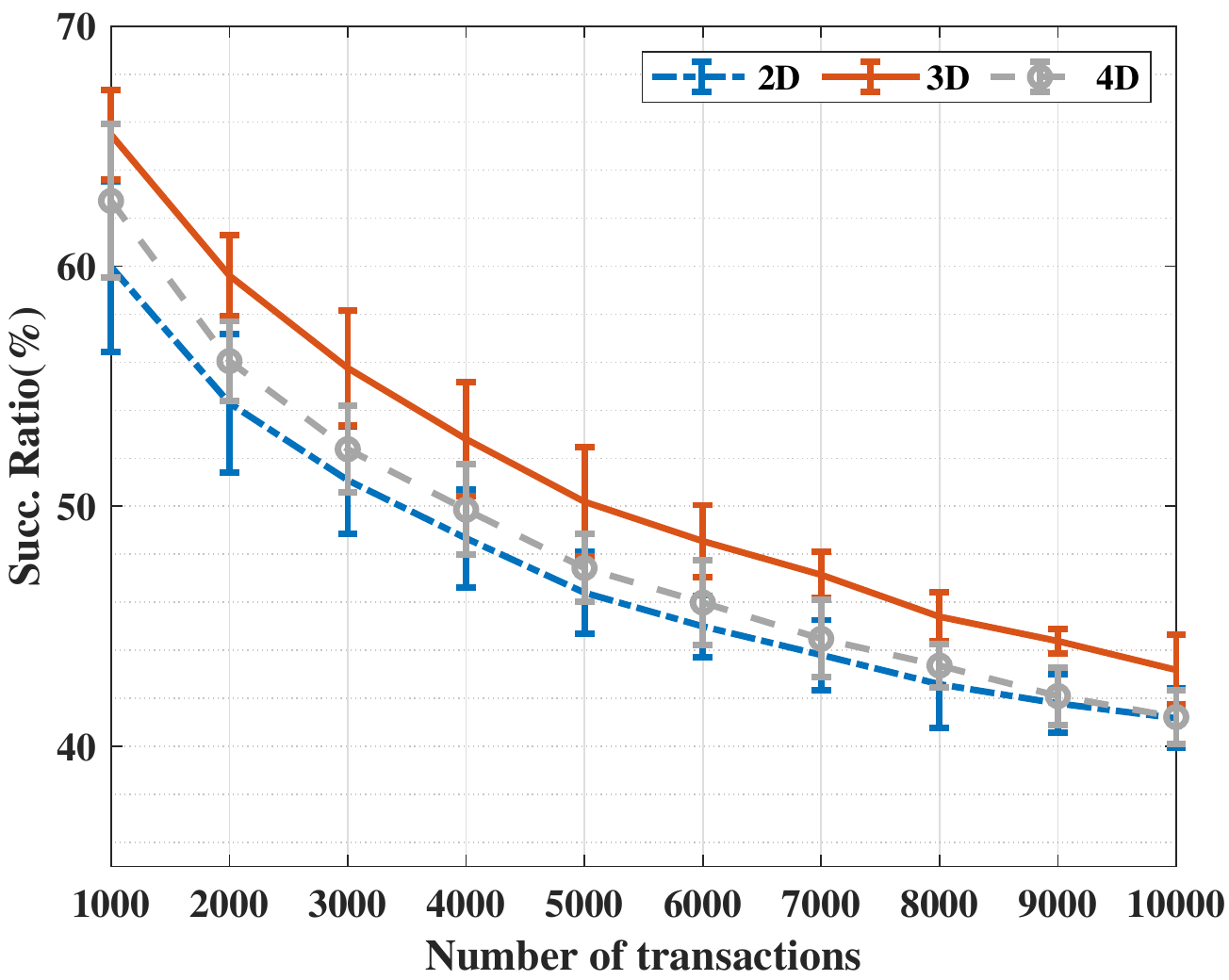}
		\vspace{-3ex}
		\caption{Routing succ. ratio in Lightning}
		\label{fig:dimension_lightning}
	\end{minipage}
	\begin{minipage}[t]{0.24\textwidth}
		\centering
		\includegraphics[width=0.98\textwidth]{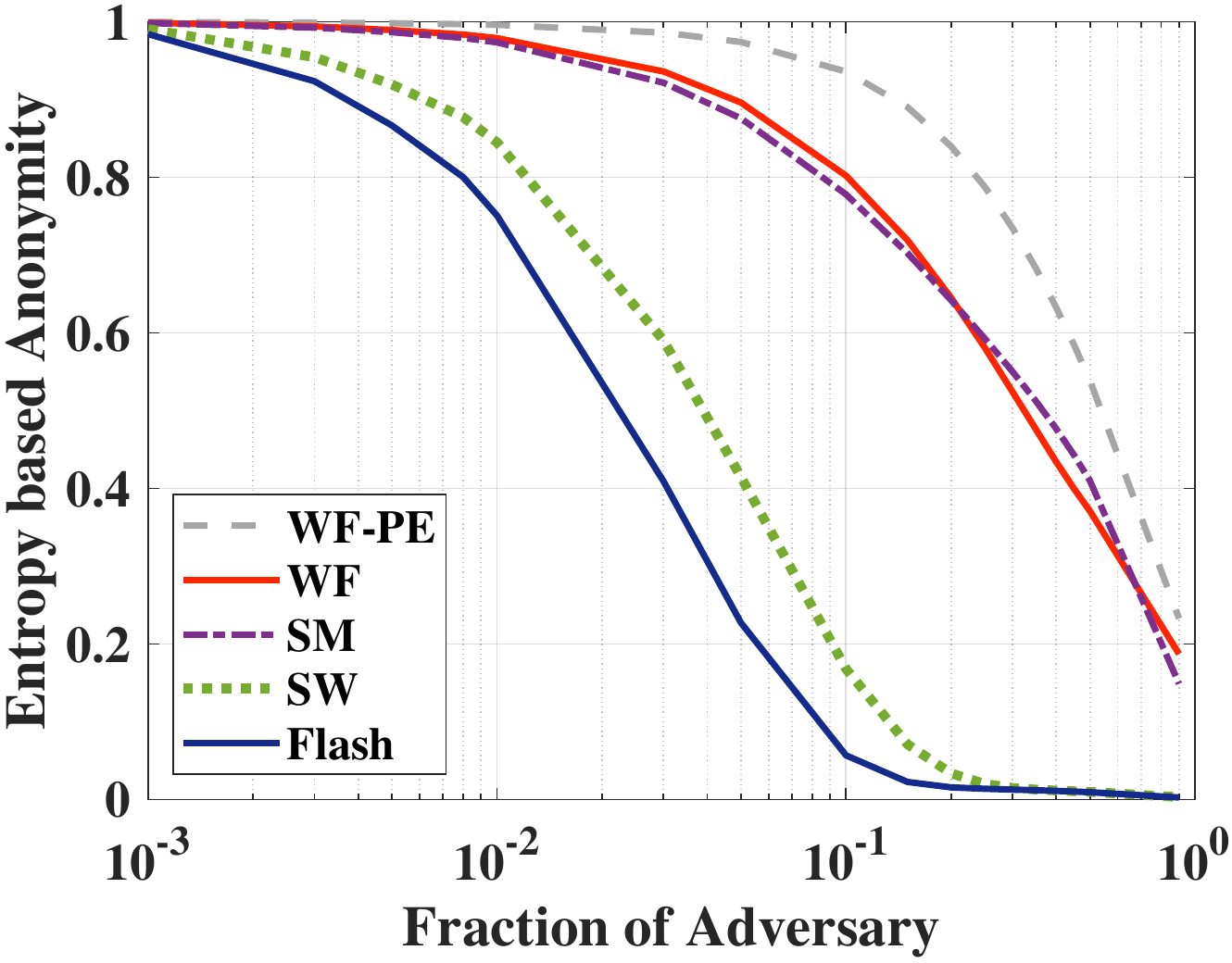}
		\vspace{-3ex}
		\caption{Anonymity of Ripple}
		\label{fig:privacy_ripple}
	\end{minipage}
	\begin{minipage}[t]{0.24\textwidth}
		\centering
		\includegraphics[width=0.98\textwidth]{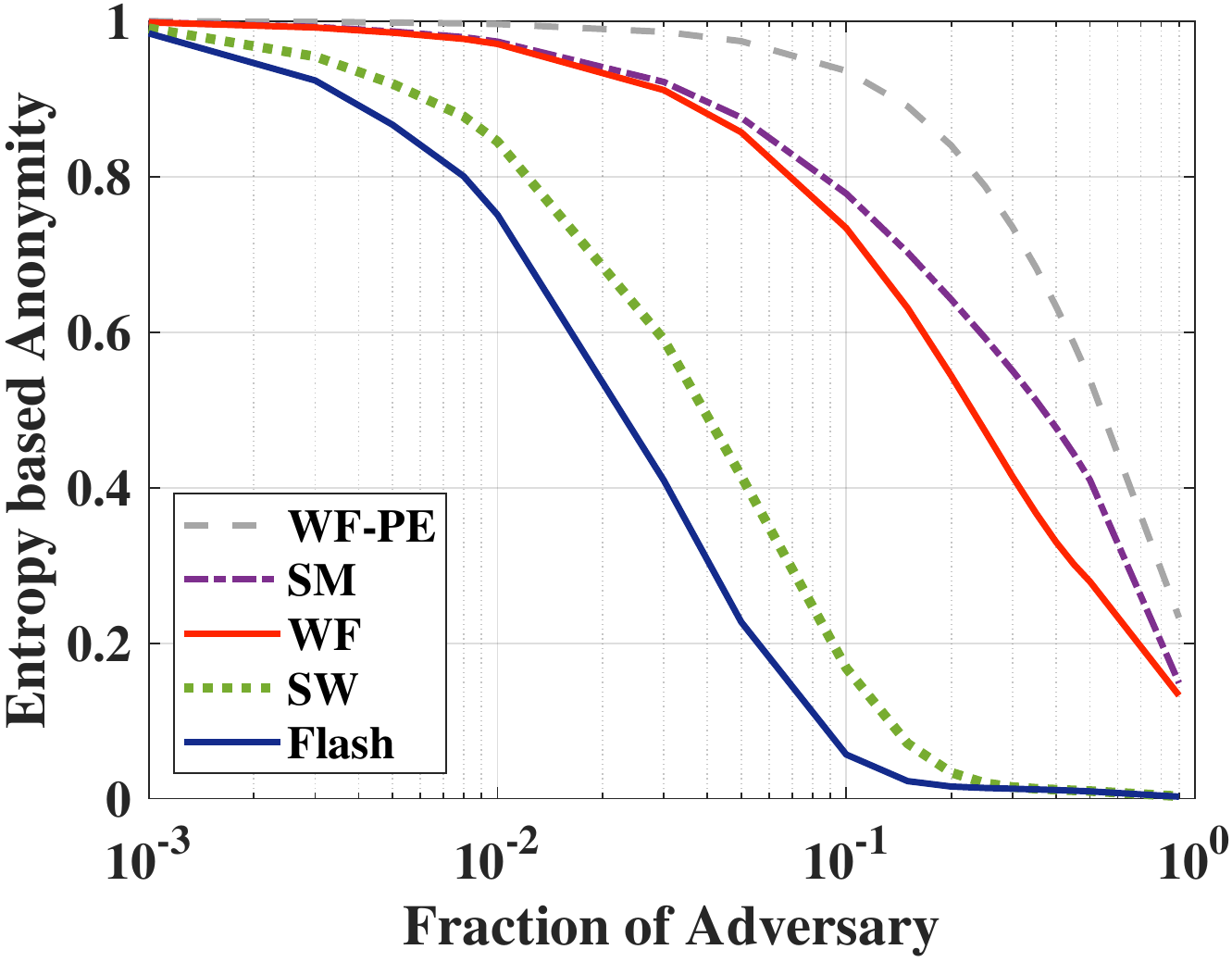}
		\vspace{-3ex}
		\caption{Anonymity of Lightning}
		\label{fig:privacy_lightning}
	\end{minipage}
\vspace{-2ex}
\end{figure*}

\textbf{Storage cost in each node.}
We now show the storage efficiency of WebFlow by comparing the average states maintained in each node. In Shortest Path, Spider and Flash, every node needs to locally store the topology of the network, including all the information of the links. Besides, in Flash, each node needs to maintain a routing table for each payment, and periodically refreshed it when the local network topology updates. 
In SilentWhispers, the landmarks need to store the network topology as well as the paths to all the nodes.
Each node only maintains the paths to all the landmarks. 
For SpeedyMurmurs, the coordinate is assigned according to the landmarks, and the length of the coordinate is depending on the depth of the node in the spanning tree. Since there are always several landmarks in the system, each node has to store several coordinates. Different from these schemes, in WebFlow, nodes only need to maintain the information of their neighbors, including the coordinates and the links to them.
Figure~\ref{fig:state} shows the average states maintained in each node. For both Ripple and Lightning, both versions of WebFlow cost less than other routing algorithms.

\textbf{Communication cost.}
We first evaluate communication cost to see if our algorithms can achieve low overhead of routing. The communication cost is computed as the total number of probing messages sent over the network.
Since SilentWhispers, SpeedyMurmurs, Shortest Path, and WebFlow-PE are static routing schemes, we consider the number of probing messages the same as their path length.
Spider and Flash use multiple paths. They first select several paths and make payments after probing the path. Moreover, Flash only probes a path when it cannot deliver the payment in full. So the number of probing messages along a path is proportional to the number of hops of the path in Spider and Flash.
MDT-based WebFlow only does probing if a node needs to reach a multi-hop DT neighbor.
In this case, the node will probe the path to its DT neighbor to see if the path has enough capacity to support the payment.
Thus the number of probing messages might be even less than the hop counts.
Figure~\ref{fig:probing} shows the comparison results with 1000 transactions. The results here demonstrate that WebFlow indeed efficiently reduce the probing message overhead compared to state of the art in all of the four topologies.



\textbf{Performance with different network load.} 
We also vary the number of transactions to test the performance of our system with different loads. 
With the increase of the number of transactions, the success ratio of all schemes decreases in both Ripple and Lightning topologies as shown in Figure~\ref{fig:dynamic_ripple} and ~\ref{fig:dynamic_lightning}. This is because as more transactions flowing into the network, more links are saturated in one direction, making them cannot be used for future transactions.
Note that in both topologies, the success ratio of the case that only considers small transactions is always higher than the case that considers all the transactions. The reason is that large transactions are easier to use up the links. Besides, large transactions are less acceptable than small ones since there are fewer links with enough capacity to support them.
In most situations, MDT-based WebFlow shows a higher success ratio and success volume compared to other methods. The only exception is for the success ratio of Lightning, where WebFlow has a lower success ratio than  Spider. 

\textbf{Choice of Dimensionality.}
We perform experiments for ripple and lightning networks embedded in 2D, 3D, and 4D virtual spaces and evaluate the performance in terms of success ratio. Figure~\ref{fig:dimension_ripple} and Figure~\ref{fig:dimension_lightning} show the results of routing performance for the two network topologies respectively. For both topologies, 3D outperforms 2D. For 4D, the results are not much better than those of 3D in ripple, and even has lower success rati in lightning. This observation is consistent with the PCA results in Fig.~\ref{fig:svd}. Besides, in a 4-dimensional space, since nodes have more DT neighbors to maintain and have to keep the coordinates of 4-tuples, both the storage and communication costs will increase compared to 2D and 3D. Hence, we choose 3D in all other experiments.

\begin{figure*}[t]
	\centering
	\begin{minipage}[t]{0.49\textwidth}
		\centering
		\includegraphics[width=0.49\textwidth]{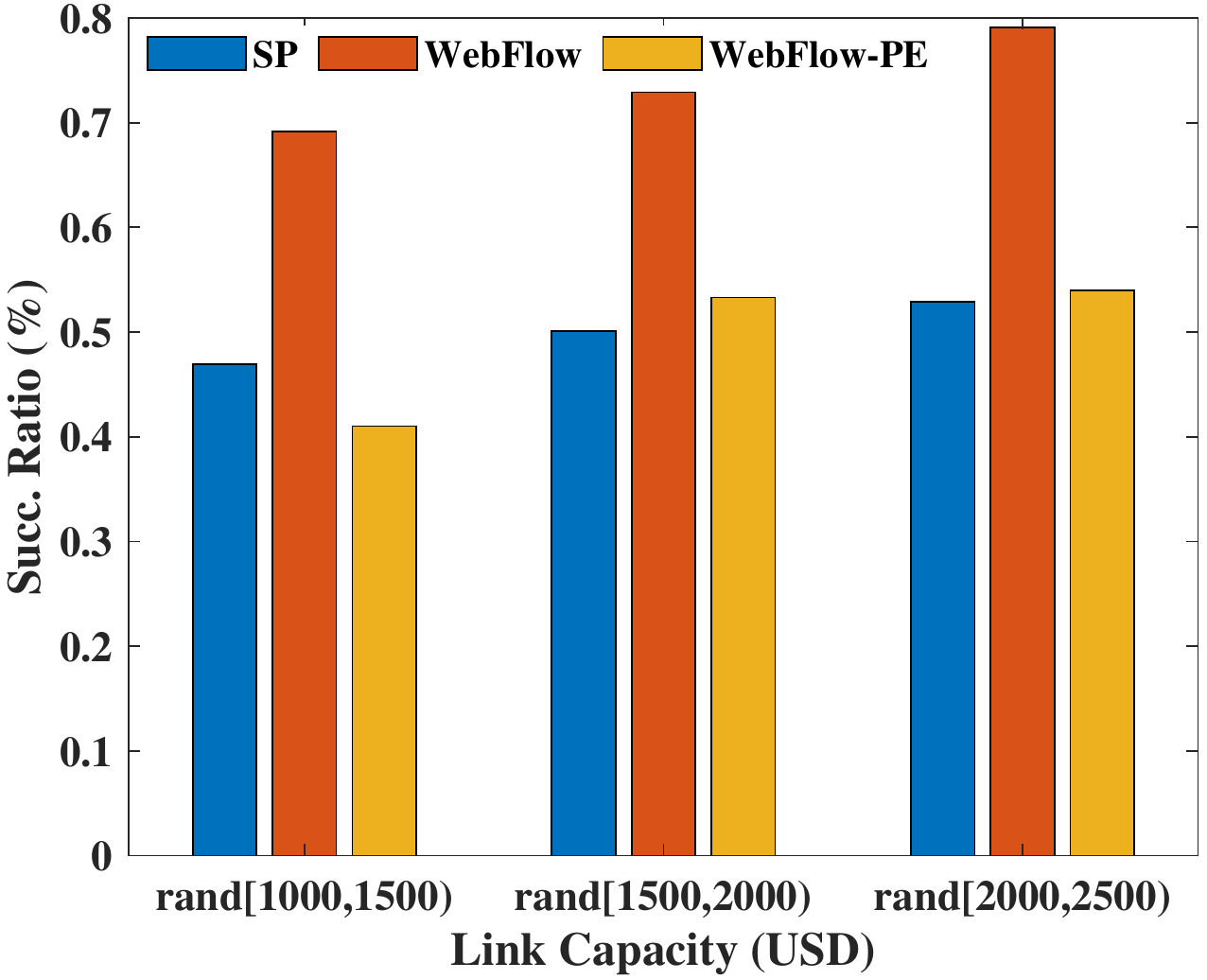}
		\includegraphics[width=0.49\textwidth]{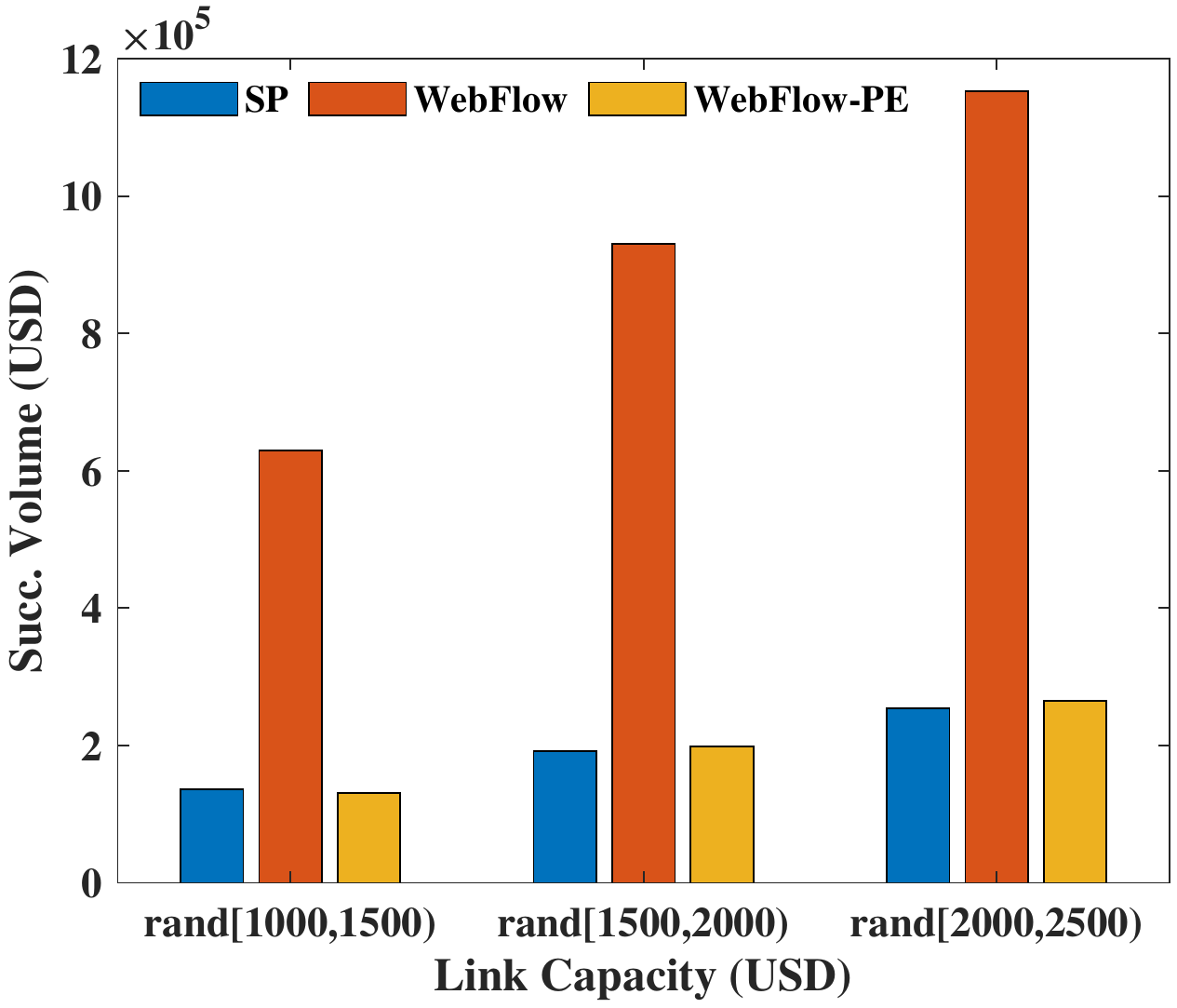}
		\vspace{-2ex}
		\caption{Testbed results of the 50-node network.}
		\label{fig:testbed_50}
	\end{minipage}
	\begin{minipage}[t]{0.49\textwidth}
		\centering
		\includegraphics[width=0.49\textwidth]{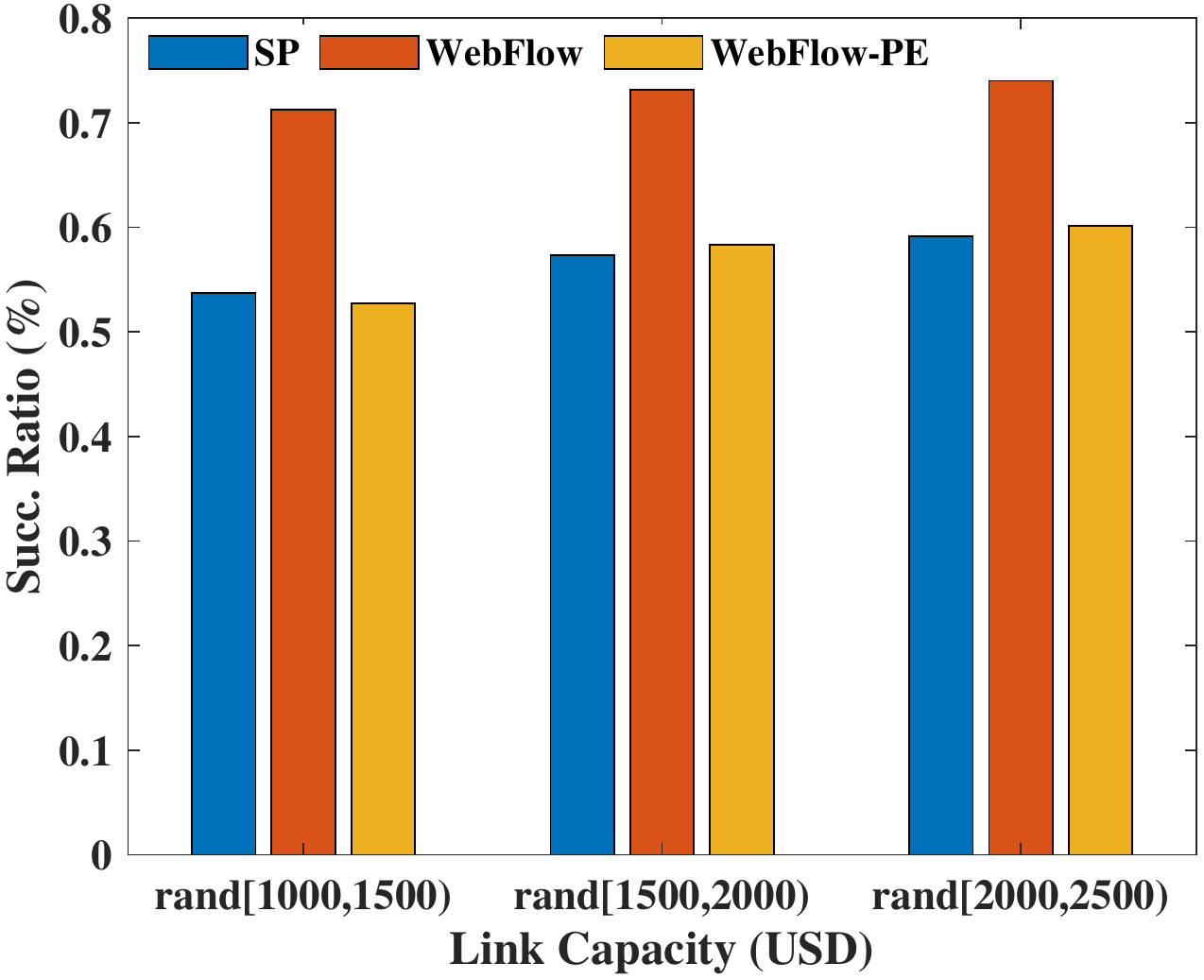}
		\includegraphics[width=0.49\textwidth]{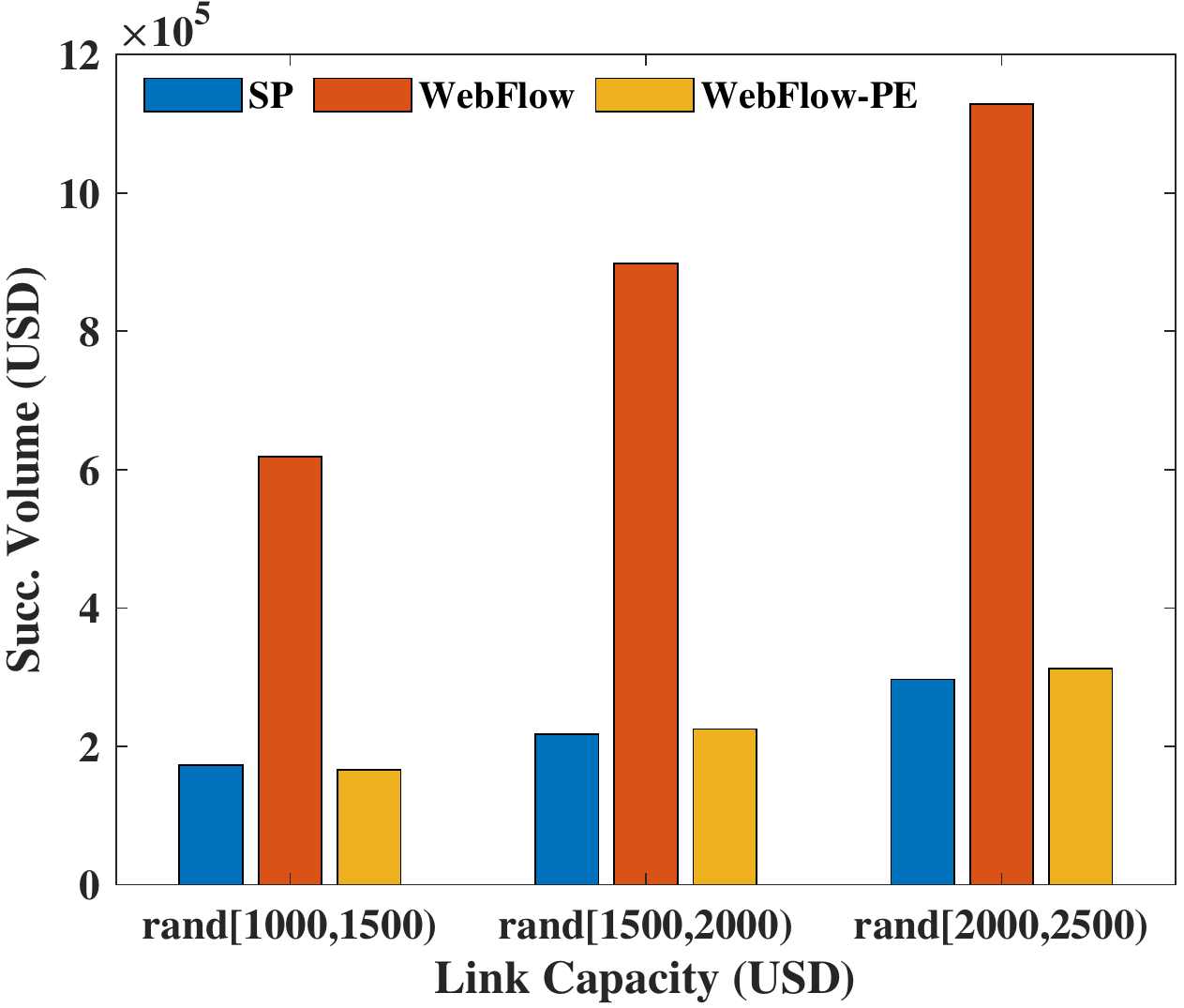}
		\vspace{-2ex}
		\caption{Testbed results of the 100-node network.}
		\label{fig:testbed_100}
	\end{minipage}
\vspace{-2ex}
\end{figure*}

\textbf{Anonymity.} 
We now demonstrate the comparison of anonymity among WebFlow and Benchmarks. WebFlow-PE shows better anony-mity measure as shown in Figs~\ref{fig:privacy_ripple} and \ref{fig:privacy_lightning}. In SilentWhispers and Flash, as long as the attacker is standing on the path, it will know exactly who the sender or recipient is. In MDT-based WebFlow, since the node needs to compute the distance between its neighbors and recipient to find the neighbor closest to the recipient, all the nodes along the path will know the information about the recipient. In SpeedyMurmurs, although it improves privacy by using anonymous return addresses, attackers can still infer that some nodes may be the sender or recipient with higher probability from the knowledge of tree constructions. 
In WebFlow-PE, nodes can only know the routing direction, instead of the coordinate of the recipient, which reduces privacy leakage. This comparing result does follow our analysis.


\subsection{Testbed Evaluation}
We conduct a testbed evaluation to further investigate WebFlow’s performance.
We implement the prototype in Golang with TCP for network communication. The prototype first generates the network topology and assigns coordinates to each node at launch time.
Upon seeing a new payment, it runs the routing algorithm and sends it out. we represent each node of the PCN as a single process running in the WebFlow prototype.
We build a PCN topology based on Waxman topology generations~\cite{waxman1988routing}. 
The routing algorithm includes three functions: distributed routing, probing, and commit. We describe the details in the following.

\textbf{Distributed Routing.} It is the most important part of our system. We implement a simple distributed routing scheme that each node is responsible for finding the next hop based on the received message. We show the message format of our prototype in Table~\ref{table:message}. 
Type field shows the message type, and also includes 1 bit indicating the scheme of routing. The destination field contains the coordinates of the recipient in the MDT-based WebFlow scheme, while shows the direction function in the WebFlow-PE scheme. Upon receiving a ROUTE message, a node can get routing direction from this field and forward it to the next hop according to the Type field. Besides, the nodes need to record TransID, last-hop, and next-hop locally for further commitment. Recipient returns ROUTE\_ACK if path found, and ROUTE\_NACK if path not found or with insufficient capacity. In the MDT-based WebFlow scheme, the nodes on the reversed path will replace the direction field with the coordinate of its last hop and forward the modified message to it. In the WebFlow-PE scheme, the recipient simply replaces the direction field with a reversed direction function.

\begin{table}[t]
	\centering
	\small
	
	{\caption{Message format in our prototype.}
		\label{table:message}
				\vspace{-2ex}
	}
	{
		\begin{tabular}{c|c}
			\hline
			Field & Description \\
			\hline
			TransID & A unique ID of a (partial) payment  \\
			Type & Message type and routing scheme  \\
			Direction & Routing direction of this message  \\
			Capacity & Probed link capacity \\
			Commit & Total funds for this payment \\
			\hline
		\end{tabular}
	}
	\vspace{-3ex}
\end{table}

\textbf{Probing.}
It is only used in the MDT-based WebFlow since the node needs to collect the ever-changing link balance to determine which link to chose for the next hop. While in WebFlow-PE, the path is determined according to the protocol. We can simply commit the payment without probing. In MDT-based WebFlow, probing only takes place when a node finds that a candidate next hop is a DT neighbor. It then probes the underlying physical path of the virtual link to check if 
it can support the payment. This node initiates probing by constructing a PROBE message for the virtual link and initiates the Capacity field to the payment value. The intermediate nodes compare their current balances with the Capacity field.
If their balances are less than the value in the Capacity field, they replace the Capacity field with their current balances. Otherwise, they do nothing. After receiving the probing message, the node simply compares the value in the Capacity field with the value in the Commit field, and determine whether this virtual link can support this payment.

\textbf{Commit.} It is the simple step to commit the payment. After finding a path for the payment, or each sub-payment, the sender sends a COMMIT message in each path. All the intermediate nodes receiving the message search for the next hop according to the TransID stored locally. Then they update their balances according to the COMMIT field and forward the message to the next hop. If an intermediate node finds that its balance is less than the payment amount, it constructs a COMMIT\_NACK message sending back to the sender. All the nodes that receive COMMIT\_NACK messages will then recover their balances.

Figure~\ref{fig:testbed_50} and Figure~\ref{fig:testbed_100} show the performance with different link capacities. The success ratio and success volume of MDT-based WebFlow are much higher than Shortest Path in both two topologies with a different number of nodes. This is because we consider link capacities when routing. The results demonstrate the effectiveness of MDT-based WebFlow which selects a good path to improve throughput. However, the performance of WebFlow-PE is similar to Shortest Path. The reason is that the ideas of these two algorithms are quite similar. We both first select a path without considering link capacities and then check if the path can satisfy the payment. The result is still acceptable since we achieve better anonymity at the cost of lower performance in WebFlow-PE.

%% file: 91-Discussion.tex
\section{Future Work}
\label{sec:future}

In this work, WebFlow achieves low storage and communication cost in dynamic PCNs. But there are several aspects we have not taken into consideration yet and may further investigate in future.

\textbf{\it Multi-path.}
Our existing design of MDT-based WebFlow only uses one path for small payments. For large payments, we treat them as several small payments and find a path for each sub-payment. It can support many payments most of the time, but in some cases, one channel may not have enough capacity to forward the payment. In WebFlow-PE, we even only use one determined path for both small and large payments.
So we can consider using multi-path for a payment stuck at the bottleneck. For MDT-based WebFlow, if a node $u$ finds that it does not have any neighbor with enough capacity to support the payment, $u$ can then divide the payment into several parts, and relies on several neighbors to continue routing. And we need to set an upper bound of multi-path to prevent too many branches of the path. In this way, we can further support more payments in the system.
For WebFlow-PE, we can also realize multi-path by selecting several intermediate nodes first before routing. Here, the intermediate nodes can be virtual nodes that do not exist. Because they only help to generate a direction function. The sender can first generate some pairs of coordinates randomly to be served as virtual intermediate nodes $v_1, v_2$ (we use 2 as an example). And it generates the direction function $\vec{l_1}$ based on sender and $v_1$,  $\vec{l_2}$ based on $v_1$ and $v_2$, $\vec{l_3}$ based on $v_2$ and recipient. Then the sender can first routing to the direction of $\vec{l_1}$, followed by $\vec{l_2}$ and $\vec{l_3}$. By selecting different groups of virtual intermediate nodes, the sender will find several paths based on different direction functions.

\textbf{\it Concurrency.}
In the design of WebFlow, we assume that payments come in sequence, and we can process them one at a time. But it is unrealistic in real-world systems. There are a large number of payments to be processed in the system and many of them may come together.
When finding paths for these concurrent payments, if several payments want to use the same channel while routing, two situations might happen. First, the two paths share this channel, but a solution needs to be found when the channel capacity is not sufficiently high. 
Second, the channel is reserved in a first-come-first-serve manner.  Then we need to provide a solution to find another path for the second transaction. 

\textbf{\it Unilateral problem.}
In PCNs, channels can become unidirectional over time, and thus block further transactions through them. This happens especially when elephant payments coming. Besides, in some other common application scenarios such as retail payments, the payment flows are always from the customers to merchants. It will easily lead to channel imbalance and seriously affected the throughput. However, in our system, we assume that all the users are equal, and payments flow in the PCN with balance. To support more application scenarios such as retail payments, we may further modify our system model, such as to treat customers and merchants differently and assume their channels have different characteristics.

%% file: 92-Conclusion.tex
\vspace{-3ex}
\section{Conclusion}
\label{sec:conclusion}

In this work, we present the design of a scalable and decentralized routing solution called WebFlow for large and dynamic PCNs. WebFlow includes two protocols: MDT-based WebFlow and WebFlow-PE. 
The first one provides a high success rate and success volume of payments. The second one, the privacy enhancement version of WebFlow, achieves destination anonymity by using routing with a distributed Voronoi diagram. Both protocols demonstrate low per-node cost and high network resource utilization.
The evaluation results using simulations and prototype implementation demonstrate that WebFlow significantly outperforms existing solutions, especially on per-node cost efficiency, while maintaining high resource utilization and success rate.


%% file: 10-appendix.tex
\section{Appendix}
\label{sec:appendix}

\subsection{Pseudocode}

\begin{algorithm}[h]
	\caption{MDT-based WebFlow Forwarding Protocol at node $u$}
	\label{Algorithm_MDT}
	
	\begin{algorithmic}[1]
		\REQUIRE A payment sent to $r$ with demand $\omega$
		\ENSURE Next hop $v$
		
		\IF{$d(u,r) == 0$}
		\RETURN $u$
		\ENDIF
		
		\STATE //Case 1: Search for direct neighbors
		\IF{there exists $v | v \in C_u$ closest to the recipient\;}
		\IF{$\psi_{uv}$ $\ge$ $\omega$}
		\RETURN $v$
		\ENDIF
		\ELSE
		\FOR{each node $v$ in $C_u$}
		\IF{$d(v,r)<d(u,r)$ and $\psi_{uv} \ge \omega$}
		\RETURN $v$
		\ENDIF
		\ENDFOR
		\ENDIF
		
		\STATE //Case 2: Search for DT neighbors
		\IF{there exists $v | v \in N_u$ closest to the recipient}
		\STATE Probe each channel of the virtual link from $u$ to $v$ to obtain their capacity $\psi_p$\;
		\STATE Find the bottleneck capacity $\omega_{min}$ = min $\psi_p$\;
		\IF{$\omega_{min}$ $\ge$ $\omega$}
		\RETURN $v$
		\ENDIF
		\ELSE
		\STATE i = 0
		\FOR{each node $v$ in $N_u$}
		\IF{$d(v,r)<d(u,r)$ and $i \le 5$}
		\STATE $i++$
		\STATE Probe each channel of the virtual link from $u$ to $v$ to obtain their capacity $\psi_p$
		\STATE Find the bottleneck capacity $\psi_{min}$ = min $\psi_p$\;
		\IF{$\psi_{min}$ $\ge$ $\omega$ }
		\RETURN $v$
		\ENDIF
		\ENDIF
		\ENDFOR
		\ENDIF
		\RETURN \o
	\end{algorithmic}
\end{algorithm}

\newpage
\begin{algorithm}[h]
	\caption{WebFlow-PE Forwarding Protocol at node $u$}
	\label{Algorithm_voronoi}
	
	\begin{algorithmic}[1]
		\REQUIRE The line with a direction $\vec{l}$ and demand $\omega$
		\ENSURE Next hop $v$
		
		\STATE // Case 1: $u$ is the recipient $r$
		\IF {$u = r$ }
		\RETURN $u$
		\ENDIF
		
		\STATE 	// Case 2: continue routing\;
		\FOR {each edges $e_i$ of $u$'s Voronoi region}
		\STATE Find $u_i$ whose Voronoi region sharing edge $e_i$
		\IF{$u_i$ is not the last hop}
		\STATE $v = u_i$
		\STATE Probe each channel of the virtual link from $u$ to $v$ to obtain their capacity $\psi_p$\;
		\STATE Find the bottleneck capacity $\omega_{min}$ = min $\psi_p$\;
		\IF{$\omega_{min}$ $\ge$ $\omega$}
		\RETURN $v$
		\ELSE
		\RETURN False
		\ENDIF
		\ENDIF
		\ENDFOR
		
	\end{algorithmic}
\end{algorithm}

\newpage
\subsection{Privacy Metric}
\textbf{Why the entropy metric is better than the straightforward metric of the probability that the attacker knows the sender or receiver.}
The entropy definition of the system is more precise than the straightforward probability definition of the probability that the attacker knows the sender or receiver. For example, let us consider sender anonymity in a network of 10000 nodes. In the first system $S_1$ with attacker $A$:
\begin{itemize}
	\item $A$ discovers the source of 10\% of payments;
	\item $A$ can limit sources of 30 \% of payments to a small
	subset of nodes, $e.g.$ 50;
	\item For the other 60 \% of payments, all nodes look
	equally likely to be a sender to $A$.
\end{itemize}

In another system $S_2$, the attacker $A$:
\begin{itemize}
	\item $A$ discovers the source of 10\% of payments;
	\item For the other 90 \% of payments, all nodes look
	equally likely to be a sender to $A$.
\end{itemize}

Using the probability that $A$ knows the sender or recipient to measure the anonymity of the system, $S_1$ and $S_2$ can achieve the same anonymity of 0.9. 
In contrast, by entropy definition, the anonymity of system $S_1$ is:

\vspace{0.08in}
AS($S_1$) = $10 \% \times 0 + 30 \% \times \frac{-50 \times \log_{2}\frac{1}{50}}{-10000 \times \log_{2}\frac{1}{10000}} + 60 \% \times 1 = 0.60$
\vspace{0.08in}

And anonymity of system $S_2$ is: 

AS($S_2$) = $10 \% \times 0 + 90 \% \times 1 = 0.90$.

Results show that $S_2$ has better anonymity than $S_1$. It true since the attacker in $S_1$ knows more information about the sender than that in $S_2$. So we know that the entropy definition provides better evaluation of the system anonymity.